\documentclass[11pt]{article}

\usepackage{a4wide}
\usepackage{color}

\usepackage[bookmarks=true,bookmarksnumbered=true,setpagesize=false]{hyperref}
\usepackage{enumerate}
\usepackage{amsmath,amssymb,amsthm}

\usepackage{textcomp,mathcomp} 

\usepackage{graphicx}
\usepackage{braket}
\usepackage{url}

\usepackage[OMLmathsfit]{isomath} 
\newcommand{\mt}{\mathsfit} 
\usepackage{mathrsfs}

\newcommand{\h}{\hat}
\newcommand{\tx}{\text}

\usepackage{fontawesome5} 
\newcommand{\fs}{\text{\faStarO}}

\usepackage{comment}

\renewcommand{\t}{\text{t}}

\let\origoverrightarrow\overrightarrow

\usepackage{mathabx}

\let\overrightarrow\origoverrightarrow

\newcommand{\lip}[1]{\left\lgroup#1\right\rgroup}

\newcommand{\bp}{\begin{pmatrix}}
\newcommand{\ep}{\end{pmatrix}}
\newcommand{\bb}{\begin{bmatrix}}
\newcommand{\eb}{\end{bmatrix}}
\newcommand{\bmat}[1]{\begin{bmatrix}#1\end{bmatrix}}

\DeclareMathOperator{\diag}{diag}

\DeclareMathOperator{\PLC}{PLC}

\newcommand{\df}{\text{d}}

\newcommand{\bs}{\boldsymbol}

\newcommand{\ub}{\underbrace}

\newcommand{\bh}[1]{\boldsymbol{\hat{#1}}}

\newcommand{\al}[1]{\begin{align}#1\end{align}}
\newcommand{\als}[1]{\begin{align*}#1\end{align*}}

\newcommand{\ab}[1]{\left|#1\right|}

\newcommand{\paren}[1]{\left(#1\right)}
\newcommand{\pn}[1]{\left(#1\right)}
\newcommand{\sqbr}[1]{\left[#1\right]}

\newcommand{\Ab}[1]{\bigl|#1\bigr|}

\newcommand{\Pn}[1]{\bigl(#1\bigr)}

\newcommand{\autospace}{%
  \mathchoice%
    {\!}
    {\!}
    {}
    {}
}
\newcommand{\fn}[1]{\autospace\paren{#1}} 
\newcommand{\Fn}[1]{\autospace\Pn{#1}} 

\newcommand{\nn}{\nonumber\\}

\newcommand{\p}{\partial}

\newcommand{\red}[1]{{\color[cmyk]{0,0.8,1,0}#1}}
\newcommand{\green}[1]{{\color[cmyk]{0.97,0,0.75,0}#1}}
\newcommand{\blue}[1]{{\color[cmyk]{1,0.5,0,0}#1}}

\newcommand{\orange}[1]{{\color[cmyk]{0,0.5,1,0}#1}}

\usepackage{fancybox}

\usepackage{amsthm}
\theoremstyle{definition}

\newcommand{\ds}{\displaystyle}

\newcommand{\ov}{\over}

\newcommand{\mc}{\mathcal}

\newcommand{\lv}[1]{\overrightarrow{#1}}

\newcommand{\nab}{\bs\nabla}

\newcommand{\ttt}{\texttt}

\newcommand{\sP}{\s_\text{P}}
\newcommand{\xP}{x_\text{P}}
\renewcommand{\P}{\text{P}}

\newcommand{\pr}{\prime}

\newcommand{\sn}{\blue{\s_n}}

\newcommand{\bx}{\orange{\bs x}}
\newcommand{\lx}{\orange{\lv x}}
\newcommand{\xz}{\orange{x^0}}

\newcommand{\s}{s}

\begin{document}
\title{
Seeing through the light cone:\\
Visualizing electromagnetic fields in special relativity
}
\author{
Daiju Nakayama,\thanks{E-mail: \tt 42.daiju@gmail.com}{}\ {}
Kin-ya Oda,\thanks{E-mail: \tt odakin@lab.twcu.ac.jp}{}\ {}
and Koichiro Yasuda\thanks{E-mail: \tt yasuda@physics.ucla.edu}\bigskip\\
\it\normalsize$^*$ e-Seikatsu Co., Ltd., 5-2-32, Tokyo 106-0047, Japan
\\
\it\normalsize$^\dagger$ Department of Information and Mathematical Sciences,\\
\it\normalsize Tokyo Woman's Christian University, Tokyo 167-8585, Japan\\
\it\normalsize$^\ddag$ Department of Physics and Astronomy, University of California, Los Angeles,\\
\it\normalsize California 90095-1547, USA\\
}
\maketitle

\begin{abstract}\noindent
The theoretical framework of electromagnetism played a foundational role in Einstein's development of special relativity. To support conceptual understanding, we present a fully special relativistic computer simulation that visualizes electromagnetic fields from the perspective of a moving observer. In this simulation, the user observes electromagnetic phenomena through their past light cone and directly experiences the Lorentz force acting at that spacetime point. The electromagnetic field is computed from the subluminal motion of point charges at the intersection of their worldlines with the observer's past light cone, ensuring causal consistency and Lorentz covariance. This approach offers an interactive and intuitive representation of relativistic electromagnetism. It provides insight into how electric and magnetic fields transform across inertial frames, and serves as a bridge between abstract formalism and physical intuition. The simulation also lends itself to pedagogical use in courses on special relativity or electrodynamics.
\end{abstract}

\newpage

\tableofcontents

\newpage

\section{Introduction}

Special relativity, together with quantum mechanics, forms a foundational pillar of modern physics, providing a framework for understanding spacetime and fundamental interactions~\cite{ParticleDataGroup:2022pth}.
Unlike quantum phenomena, which are beyond the reach of our everyday classical perception, relativistic effects such as time dilation and length contraction can be directly observed, making them particularly suitable for visualization-based learning. Interactive simulations offer an intuitive way to approach these abstract concepts.

Students often face significant conceptual difficulties in mastering special relativity, particularly when grappling with frames of reference, simultaneity, and transformation laws~\cite{mcdermott1999resource, scherr2001student}. Traditional instruction methods, which primarily rely on algebraic derivations and static spacetime diagrams, frequently fall short in promoting deep understanding. In response, educators have developed a variety of innovative tools---including interactive simulations and multimedia platforms---to dynamically visualize relativistic phenomena~\cite{wieman2008phet, mcgrath2010student, kortemeyer2019game}; see also Ref.~\cite{Arriassecq2007} for a collection of related work. These efforts suggest that combining interactive visualizations with guided conceptual inquiry can significantly enhance learning outcomes.

Previous computational work has primarily focused on relativistic kinematics, offering simulations of time dilation, length contraction, and frame transformations~\cite{wieman2008phet, mcgrath2010student, kortemeyer2019game}. These efforts have significantly aided conceptual understanding, but have largely overlooked the role of electromagnetism---even though Einstein's formulation of special relativity was originally motivated by the transformation properties of electric and magnetic fields~\cite{Einstein:1905ve}.

In Ref.~\cite{10.1093/ptep/ptx127}, we introduced a visualization method based on the past-light-cone (PLC) perspective, enabling a Lorentz-covariant treatment of relativistic motion. In this paper, we extend that approach to include electromagnetic fields, providing, to our knowledge, the first fully relativistic and Lorentz-covariant computer simulation of electromagnetic fields from the viewpoint of a moving observer.

In our simulation, users perceive the world through intersections with their PLC, and experience electromagnetic fields in a manifestly Lorentz-covariant manner. This builds on a theoretical framework that formulates electromagnetic fields in terms of PLCs~\cite{Nakayama:2024wfi}, enabling a consistent and causally grounded approach to relativistic electrodynamics.

Our implementation consists of three main components:
\begin{enumerate}
    \item At each spacetime point on the user's PLC, the electromagnetic field tensor is evaluated and transformed into the user's rest frame, demonstrating the frame-dependent mixing of electric and magnetic components.
    \item Every charged particle, including the user, experiences the Lorentz force based on the electromagnetic field at its own spacetime location.
    \item The electromagnetic field at any spacetime point is computed from the subluminal motion of point charges at the intersection of their worldlines with the PLC of that point, ensuring causal consistency and relativistic covariance.
\end{enumerate}

By combining these elements into a single interactive simulation, we provide an intuitive and visually coherent representation of relativistic electromagnetism. The simulation operates in real time as the observer moves, serving as a bridge between abstract formalism and physical intuition. It also serves as a pedagogical demonstration for courses on special relativity or electrodynamics.

This paper is organized as follows.
In Sec.~\ref{sec:Covariant kinematics for visualization}, we describe the Lorentz-covariant kinematics underlying the simulation.
In Sec.~\ref{sec:Drawing the world on the past light cone}, we describe how the visible world is constructed from intersections with the PLC.
In Sec.~\ref{elemag review section}, we present a covariant description of charged particle motion and introduce the electromagnetic potential in a relativistic setting.
In Sec.~\ref{covariant formalism section}, we apply the Lorentz-covariant formalism to compute electromagnetic fields from moving charges.
Sec.~\ref{Implementation} describes the practical aspects of implementing these theoretical concepts, detailing the computational framework and visualization techniques.
Finally, Sec.~\ref{Conclusion} summarizes the key contributions of this work.

\section{Covariant kinematics for visualization}\label{sec:Covariant kinematics for visualization}

This section sets up the Lorentz-covariant framework needed to describe particle motion in special relativity, building on the formalism introduced in Ref.~\cite{10.1093/ptep/ptx127} and restructuring key elements to suit our simulation-oriented perspective. Coordinates, proper time, velocity, and acceleration are introduced, along with the role of PLCs in determining what is observable. These concepts lay the foundation for frame-independent, causally consistent simulations.

We base our formulation on the following three principles of special relativity:
\begin{enumerate}[I.]
\item The constancy of the speed of light in vacuum.\label{constant speed of light}
\item Lorentz invariance, implying that the laws of physics are the same for all inertial observers.\label{Lorentz invariance}
\item The equivalence principle, adapted for specific contexts within special relativity.\label{equivalence principle}
\end{enumerate}

These principles guide the structure of this section: principle~\ref{constant speed of light} motivates the formulation in Sec.~\ref{spacetime coordinate section}; principle~\ref{Lorentz invariance} underpins the developments in Secs.~\ref{Lorentz transformation}--\ref{rest frame section}; and the adapted equivalence principle~\ref{equivalence principle} finds its application in Sec.~\ref{eom section}.

\subsection{Spacetime coordinates}\label{spacetime coordinate section}
We first prepare a Cartesian coordinate system for a reference frame, which we call the \emph{world frame} hereafter:
\al{
\lv x
	&=	\pn{x^0,\bs x}
	=\pn{x^0,x^1,x^2,x^3},
}
where
\al{
x^0:=ct
	\label{x0 defined}
}
denotes the time $t$ multiplied by the (constant) speed of light $c$ and $\bs x=\pn{x^1,x^2,x^3}$ is the spatial coordinates, which are more frequently (but not in this paper) written as $\pn{x,y,z}$.
Hereafter, we call $x^0$, having the dimension of length, the \emph{time length}.
If one does not intend to change the speed of light depending on the game/simulation settings, one may take the \emph{natural units}
\al{
c=1.
}
Throughout this paper, we keep the non-natural units for the possible reader's ease.

We also employ the matrix notation
\al{
\lv x
	&=	\bmat{x^0\\ \bs x}=\bmat{x^0\\ x^1\\ x^2\\ x^3},&
\lv x^\t
	&=	\bmat{x^0&\bs x^\t}=\bmat{x^0&x^1&x^2&x^3},
}
where the superscript ``t'' denotes the transpose. 

In the actual implementation in Sec.~\ref{Implementation}, the spatial and time-length coordinates are indexed as
\al{
\bmat{\tt x\\ \tt y\\ \tt z\\ \tt w}
	=	\bmat{\texttt{x[0]}\\ \texttt{x[1]}\\ \texttt{x[2]}\\ \texttt{x[3]}}
	&:=	\bmat{x^1\\ x^2\\ x^3\\ x^0}.
}
The authors are sorry for the confusing notation but this is how the physics and information-technology communities translate.

For spatial vectors $\bs x=\pn{x^1,x^2,x^3}$ and $\bs y=\pn{y^1,y^2,y^3}$, we write their inner product
\al{
\bs x\cdot\bs y
	&:=	\sum_{i=1}^3x^iy^i
	=\bmat{x^1&x^2&x^3}\bmat{y^1\\ y^2\\ y^3}
	=	\bs x^\t\bs y,\\
\bs x^2
	&:=	\bs x\cdot\bs x
	=	\pn{x^1}^2+\pn{x^2}^2+\pn{x^3}^2
	=	\sum_{i=1}^3\pn{x^i}^2,
		\label{x sqaured}\\
\ab{\bs x}
	&:=	\sqrt{\bs x^2}
	=	\sqrt{\pn{x^1}^2+\pn{x^2}^2+\pn{x^3}^2}
	=	\sqrt{\sum_{i=1}^3\pn{x^i}^2}.
}
Throughout this paper, the roman and greek letters $i,j,\dots$ and $\mu,\nu,\dots$ run for $1,2,3$ and $0,\dots,3$, respectively.
A hat symbol denotes a unit vector:
\al{
\bh x
	&:=	{\bs x\ov\ab{\bs x}}.\label{hat notation}
}

For later use, we define a matrix called \emph{metric}:\footnote{
Here and hereafter, we use the sans-serif italic font to denote matrices in the $d+1$ spacetime dimensions.
}
\al{
\mt\eta
	=	\bmat{\eta_{\mu\nu}}_{\mu,\nu=0,\dots,3}
	=	\bmat{
			\eta_{00}&\eta_{01}&\eta_{02}&\eta_{03}\\
			\eta_{10}&\eta_{11}&\eta_{12}&\eta_{13}\\
			\eta_{20}&\eta_{21}&\eta_{22}&\eta_{23}\\
			\eta_{30}&\eta_{31}&\eta_{32}&\eta_{33}
			}
	&:=	\bmat{
			-1&0&0&0\\
			0&1&0&0\\
			0&0&1&0\\
			0&0&0&1
			}.
			\label{eta defined}
}
Since $\mt\eta^2=\mt I$, with $\mt I$ being the identity matrix, the inverse of $\mt\eta$ becomes identical to itself: $\mt\eta^{-1}=\mt\eta$, where
\al{
\mt\eta^{-1}
	=	\bmat{\eta^{\mu\nu}}_{\mu,\nu=0,\dots,3}
	=	\bmat{
			\eta^{00}&\eta^{01}&\eta^{02}&\eta^{03}\\
			\eta^{10}&\eta^{11}&\eta^{12}&\eta^{13}\\
			\eta^{20}&\eta^{21}&\eta^{22}&\eta^{23}\\
			\eta^{30}&\eta^{31}&\eta^{32}&\eta^{33}
			}
	=	\bmat{
			-1&0&0&0\\
			0&1&0&0\\
			0&0&1&0\\
			0&0&0&1
			}.
}
Using $\mt\eta$, we define the Lorentzian inner product and the Lorentzian (squared-)norm:
\al{
\lip{\lv x,\lv y}
	:=	\lv x^\t\mt\eta\lv y
	&=	\bmat{x^0&x^1&x^2&x^3}\bmat{-1&0&0&0\\ 0&1&0&0\\ 0&0&1&0\\ 0&0&0&1}\bmat{y^0\\ y^1\\ y^2\\ y^3}\nn
	&=	-x^0y^0+\bs x\cdot\bs y
	=	-x^0y^0+x^1y^1+x^2y^2+x^3y^3,\\
\lip{\lv x}^2
	:=	\lip{\lv x,\lv x}
	&=	-\pn{x^0}^2+\bs x^2
	=	-\pn{x^0}^2+\pn{x^1}^2+\pn{x^2}^2+\pn{x^3}^2.
}
We might sometimes write
\al{
\lv x\cdot\lv y
	&:=	\lip{\lv x,\lv y},&
\lv x^2
	&:=	\lip{\lv x}^2.
	\label{ordinary notation}
}

\subsection{Lorentz transformations}\label{Lorentz transformation}
Special relativity postulates the light-speed invariance.
The coordinates $\bs x=\pn{x^1,x^2,x^3}$ of a spherical light wave originating from $\bs x=0$ at the time length $x^0=0$ are prescribed by the following constraint at a time length $x^0$:
\al{
\bs x^2
	&=	\pn{x^0}^2.
}
This condition for the spherical light wave can be written as
\al{
\lip{\lv x}^2
	&=	0.\label{spherical wave}
}

In the spirit of the light-speed invariance,
a linear spacetime-coordinate transformation (represented by a matrix) $\mt\Lambda$,
\al{
\lv x
	&\to
		\lv x'
	=	\mt\Lambda\lv x,&
\bmat{x^0\\ x^1\\ x^2\\ x^3}
	&\to
		\bmat{x^{\pr0}\\ x^{\pr1}\\ x^{\pr2}\\ x^{\pr3}}
	=	\bmat{
			\Lambda^0{}_0&\Lambda^0{}_1&\Lambda^0{}_2&\Lambda^0{}_3\\
			\Lambda^1{}_0&\Lambda^1{}_1&\Lambda^1{}_2&\Lambda^1{}_3\\
			\Lambda^2{}_0&\Lambda^2{}_1&\Lambda^2{}_2&\Lambda^2{}_3\\
			\Lambda^3{}_0&\Lambda^3{}_1&\Lambda^3{}_2&\Lambda^3{}_3
			}
		\bmat{x^0\\ x^1\\ x^2\\ x^3},
		\label{Lorentz transformation on x}
}
is called the \emph{Lorentz transformation} when it leaves the left-hand side of Eq.~\eqref{spherical wave} invariant: For all $\lv x$,
\al{
\lip{\lv x'}^2
	&=	\lip{\lv x}^2&
&\Longleftrightarrow&
\pn{\mt\Lambda\lv x}^\t\mt\eta\pn{\mt\Lambda\lv x}
	&=	{\lv x}^\t\mt\eta\lv x&
&\Longleftrightarrow&
{\lv x}^\t\pn{\mt\Lambda^\t\mt\eta\mt\Lambda}\lv x
	&=	{\lv x}^\t\mt\eta\lv x.
}
That is, a transformation (represented by) $\mt\Lambda$ is called the Lorentz transformation when and only when\footnote{\label{proper orthochronous}
This relation results in $\det\mt\Lambda=\pm1$ and $\ab{\Lambda^0{}_0}\geq1$.
Hereafter, when we say Lorentz transformation, it denotes the proper ($\det\mt\Lambda=1$) orthochronous  ($\Lambda^0{}_0\geq1$) Lorentz transformation, barring the parity transformation (or space inversion) $\mt P:=\diag\fn{1,-1,-1,-1}$ and the time reversal $\mt T:=\diag\fn{-1,1,1,1}$, where ``diag'' denotes the diagonal matrix.
}
\al{
\mt\Lambda^\t\,\mt\eta\,\mt\Lambda
	&=	\mt\eta.
	\label{Lambda as Lorentz transformation}
}
This can be equivalently written as
\al{
\mt\Lambda^{-1}
	&=	\mt\eta\,\mt\Lambda^\t\,\mt\eta.
	\label{Lambda inverse explicitly written}
}

When a ``spacetime vector'' $\lv V=\pn{V^0,V^1,V^2,V^3}$ transforms the same as the coordinates~$\lv x$ under the Lorentz transformations~\eqref{Lorentz transformation on x}, namely $\lv V\to\lv V'=\mt\Lambda\lv V$, we call $\lv V$ a \emph{covariant} vector.\footnote{
In this paper, we call both the contravariant and covariant vectors the covariant vectors in the spirit that both are Lorentz covariant.
}
Special relativity is a theory that describes everything in terms of covariant and invariant quantities.

\subsection{Light cones}
\begin{figure}[tp]
\centering
\includegraphics[width=0.8\textwidth]{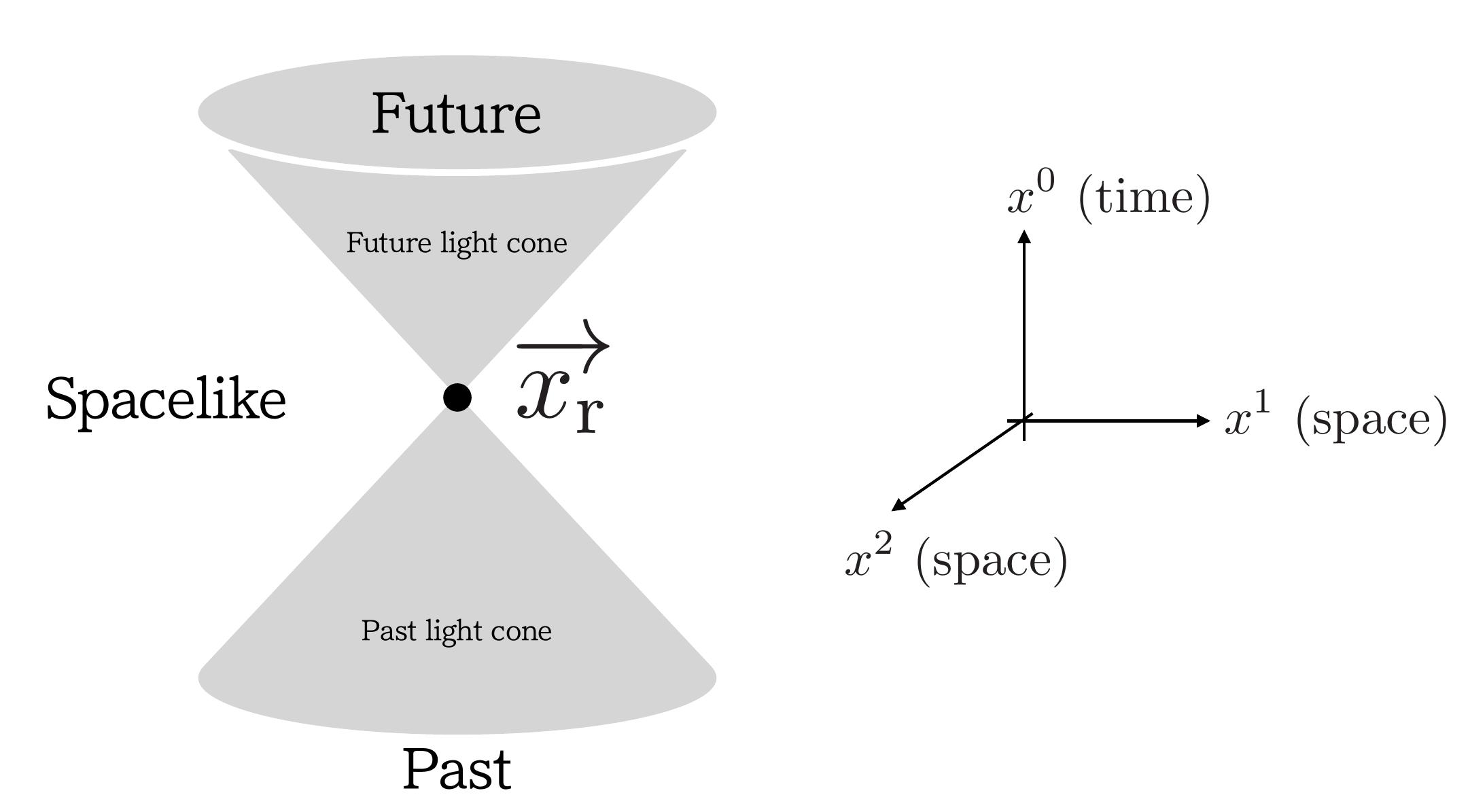}
\caption{Schematic figure (in two spatial dimensions) for the past, spacelike, and future regions of $\protect\lv{x_\tx r}$, as well as its future and past light cones.\label{lightcone figure}}
\end{figure}

The above argument sets the spacetime origin as the reference point. Now we generalize it.
Given a reference point $\lv{x_\tx r}$ and another arbitrary point $\lv x$, let us consider the following Lorentz \emph{invariant}:
\al{
\lip{\lv x-\lv{x_\tx r}}^2
	&=	-\pn{x^0-x_\tx r^0}^2+\pn{\bs x-\bs x_\tx r}^2.
}
With it, we may divide the whole spacetime into three regions (see Fig.~\ref{lightcone figure}):
\begin{itemize}
\item $\lv x$ belongs to the \emph{past} region of $\lv{x_\tx r}$ if the separation is \emph{timelike} $\lip{\lv x-\lv{x_\tx r}}^2<0$ and if $x^0<x_\tx r^0$.
\item $\lv x$ belongs to the \emph{spacelike} region of $\lv{x_\tx r}$ if the separation is \emph{spacelike} $\lip{\lv x-\lv{x_\tx r}}^2>0$.
\item $\lv x$ belongs to the \emph{future} region of $\lv{x_\tx r}$ if the separation is timelike $\lip{\lv x-\lv{x_\tx r}}^2<0$ and if $x^0>x_\tx r^0$.
\end{itemize}
These three regions are separated by the \emph{light cones}:
\begin{itemize}
\item $\lv x$ belongs to the \emph{past light cone} (PLC) of $\lv{x_\tx r}$ if the separation is \emph{lightlike} $\lip{\lv x-\lv{x_\tx r}}^2=0$ and if $x^0<x_\tx r^0$.
\item $\lv x$ belongs to the \emph{future light cone} of $\lv{x_\tx r}$ if the separation is \emph{lightlike} $\lip{\lv x-\lv{x_\tx r}}^2=0$ and if $x^0>x_\tx r^0$.
\end{itemize}
We emphasize that the three regions and the light cones are defined independently of the chosen coordinate system.\footnote{\label{time ordering}
The order of times is Lorentz invariant when the separation is either timelike or lightlike; see also footnote~\ref{proper orthochronous}.
}

\subsection{Proper time and covariant velocity}
Let us consider objects $O_1,O_2,\dots$. 
We write the $O_n$'s spacetime position $\lv{x_n}$.
In the next iteration of the computer program, it moves to $\lv{x_n}'$.\footnote{
It should be understood that the prime symbol here, ``$\prime$'', has nothing to do with the Lorentz transformation~\eqref{Lorentz transformation on x}.
}
From the displacement vector
\al{
\lv{\Delta x_n}
	&:=	\lv{x_n}'-\lv{x_n},&
\bmat{\Delta x_n^0\\ \Delta\bs x_n}
	&=	\bmat{x_n^{\pr0}-x_n^0\\ \bs x_n'-\bs x_n},
}
we may define a Lorentz invariant: $\lip{\lv{\Delta x_n}}^2=-\pn{\Delta x_n^0}^2+\pn{\Delta\bs x_n}^2$. A particle is called \emph{massive}, \emph{massless}, and tachyon when it moves with this Lorentz-invariant quantity being always negative, zero, and positive, respectively:
\al{
\lip{\lv{\Delta x_n}}^2
	\begin{cases}
	<0	&	\tx{massive (moving timelike)},\\
	=0	&	\tx{massless (moving lightlike)},\\
	>0	&	\tx{tachyon (moving spacelike)}.
	\end{cases}
}
The light is massless.
Hereafter, we assume all the objects (other than light rays) are massive unless otherwise stated.
We also assume that an object moves forward in the future direction: $\Delta x_n^0>0$.\footnote{
The order of time is well-defined both for massive and massless particles that move into the future and the future-light cone, respectively; see footnote~\ref{time ordering}.
We also note that there is no need to consider an object moving backward in time because, in quantum field theory, a particle moving backward in time is equivalent to an anti-particle moving forward in time, and vice versa.
}

Note that by the definition of massiveness,
\al{
\lip{\lv{\Delta x_n}}^2<0\quad\Longleftrightarrow\quad
\pn{\Delta\bs x_n}^2<\pn{\Delta x_n^0}^2,
}
the (non-covariant) velocity of the massive particle\footnote{
The expression $\bs v_n={\df\bs x_n\ov\df t}$ in the world frame might be more familiar for some readers; recall Eq.~\eqref{x0 defined}.
}
\al{
\bs v_n
	&:=	c{\df\bs x_n\ov\df x_n^0}
	=	\lim_{\Delta x_n^0\to0}c{\Delta\bs x_n\ov\Delta x_n^0}
		\label{non-covariant velocity given}
}
always satisfies
\al{
\ab{\bs v_n}<c.
	\label{slower than speed of light}
}
That is, all the massive objects move slower than the speed of light.

For each iteration, we define what we call a \emph{proper time distance} for $O_n$ by
\al{
\Delta\s_n
	&:=	\sqrt{-\lip{\lv{\Delta x_n}}^2}
	=	\Delta x_n^0\sqrt{1-\pn{\Delta\bs x_n\ov\Delta x_n^0}^2}.
	\label{proper time length defined}
}
By adding the proper time distance of all the iterations (up to the spacetime point of interest), we obtain what we call the \emph{proper time length} $s_n$ that parametrizes each world line of $O_n$;
it becomes the same as the \emph{proper time} $\tau_n:=s_n/c$ in natural units $c=1$.\footnote{
As we are in the non-natural units, we unconventionally call the proper time (multiplied by $c$) the proper time length.
}

In the limit of infinitesimal iterations, the proper time length $\s_n$ becomes a continuous parameter.
It is important that the proper time length can parametrize the $O_n$'s \emph{worldline}~$\mc W_n$ in a Lorentz-\emph{invariant} fashion, where $\mc W_n$ is the trajectory of $O_n$ in the spacetime:
\al{
\mc W_n
	&:=	\Set{\lv{x_n}\fn{\s_n}|\s_n^\tx{start}\leq\s_n\leq\s_n^\tx{cease}},
}
with $\s_n^\tx{start}$ and $\s_n^\tx{cease}$ being the proper time lengths at which $O_n$ starts and ceases to exist, respectively.

In the limit of infinitesimal lapse of the proper time length $\Delta s_n\to0$, we define the dimensionless \emph{covariant velocity}:
\al{
\lv{u_n}
	&:=	\lim_{\Delta s_n\to0}{\lv{\Delta x_n}\ov\Delta\s_n}
	=	{\df\lv{x_n}\ov\df\s_n },
	\label{covariant velocity defined}
}
where the particle position (on its worldline) is regarded as a function of its proper time length: $\lv{x_n}\fn{\s_n}$.
Hereafter, when we simply write \emph{velocity}, it denotes the dimensionless covariant velocity, and we always call $\bs v_n$ the non-covariant velocity.

By definition~\eqref{proper time length defined}, we see that
\al{
\lip{\lv{u_n}}^2
	&=	-1,&
u_n^0
	&=	\sqrt{1+\bs u_n^2}.
	\label{normalization of u}
}
It is important that $\lv{u_n}$ has only 3 independent components $\bs u_n=\pn{u_n^1,u_n^2,u_n^3}$ and $u^0$ is always given in terms of them.

In the limit of infinitesimal time-lapse, the proper time distance becomes
\al{
\df\s_n =\df x_n^0\sqrt{1-{\bs v_n^2\ov c^2}}.
}
Therefore, we obtain,
\al{
{\bs v_n\ov c}
	&=	{\bs u_n\ov\sqrt{1+\bs u_n^2}}
	=	{\bs u_n\ov u_n^0},&
\bs u_n
	&=	{{\bs v_n\ov c}\ov\sqrt{1-{\bs v_n^2\ov c^2}}}.
		\label{v and u}
}
The range of velocity is $0\leq\ab{\bs u_n}<\infty$ for $0\leq\ab{\bs v_n}<c$.

For any velocity $\lv u$, its time component
\al{
u^0=\sqrt{1+\bs u^2}\
	\pn{=	{1\ov\sqrt{1-{\bs v^2\ov c^2}}}}
	\label{gamma factor}
}
takes the values $1\leq u^0<\infty$ for the above range of velocities.
It governs the time dilation in the sense that
\al{
\df x_n^0
	&=	u_n^0\,\df\s_n,
	\label{derivative relation}
}
namely $\df t_n=u_n^0\,\df\tau_n$; see the next subsection.\footnote{
The time component $u^0=\sqrt{1+\bs u^2}$ is sometimes denoted by $\gamma\fn{\bs u}$ and is called the gamma factor.
}

\subsection{Covariant acceleration}\label{acceleration section}
Along with the covariant velocity, we define the \emph{covariant acceleration} of $O_n$ as
\al{
\lv{\alpha_n}\fn{\s_n}
	&:=	{\df\lv{u_n}\fn{\s_n} \ov \df\s_n}.
	\label{covariant acceleration defined}
}
Note that this quantity has the dimension of inverse length.

Taking the derivative of the normalization condition $\lip{\lv{u_n}}^2 = -1$ with respect to $\s_n$, we obtain
\al{
\lip{\lv{\alpha_n}\fn{\s_n}, \lv{u_n}\fn{\s_n}} = 0,
}
which shows that the time component of the covariant acceleration is not independent of the spatial ones:
\al{
\alpha_n^0
	&=	{\bs\alpha_n \cdot \bs u_n \ov \sqrt{1 + \bs u_n^2}}.
	\label{alpha0}
}
This relation holds at any proper time $\s_n$ in any inertial frame.
In particular, $\alpha_n^0 = 0$ whenever $\bs u_n = 0$.

\subsection{Transformation to instantaneous rest frame}\label{rest frame section}

For each object $O_n$ at any proper time length $\s_n$, there exists a unique reference frame in which it is momentarily at rest. This frame, known as the \emph{instantaneous rest frame}, corresponds to the viewpoint of $O_n$ itself at that moment---it is the frame with respect to which $O_n$ observes the world around it.

This frame is related to the world frame by a Lorentz transformation~\eqref{Lorentz transformation on x} determined by $O_n$'s covariant velocity $\lv{u_n}(\s_n)$, and is explicitly given by $\mt\Lambda = \mt L\Fn{\bs u_n\fn{\s_n}}$, where
\al{
\mt L\fn{\bs u}
	&:=	\bmat{u^0&-\bs u^\t\\-\bs u&I+\pn{u^0-1}\bh u\bh u^\t}
	=	\bmat{
			u^0&-u^1&-u^2&-u^3\\
			-u^1&\ds1+\pn{u^0-1}{u^1u^1\ov\ab{\bs u}^2}&\ds\pn{u^0-1}{u^1u^2\ov\ab{\bs u}^2}&\ds\pn{u^0-1}{u^1u^3\ov\ab{\bs u}^2}\smallskip\\
			-u^2&\ds\pn{u^0-1}{u^2u^1\ov\ab{\bs u}^2}&\ds1+\pn{u^0-1}{u^2u^2\ov\ab{\bs u}^2}&\ds\pn{u^0-1}{u^2u^3\ov\ab{\bs u}^2}\smallskip\\
			-u^3&\ds\pn{u^0-1}{u^3u^1\ov\ab{\bs u}^2}&\ds\pn{u^0-1}{u^3u^2\ov\ab{\bs u}^2}&\ds1+\pn{u^0-1}{u^3u^3\ov\ab{\bs u}^2}
			},
			\label{to rest frame}
}
with $u^0$ defined in Eq.~\eqref{gamma factor}. It is straightforward to check that $\mt L$ satisfies the Lorentz condition~\eqref{Lambda as Lorentz transformation} and obeys
\al{
\mt L\fn{\bs u}\lv{u}
	&=	\bmat{1\\\bs 0},&
\Pn{\mt L\fn{\bs u}}^{-1}
	&=	\mt L\fn{-\bs u}.
	\label{L inverse}
}

Hereafter, we use uppercase letters for quantities in the instantaneous rest frame of an object:
\al{
\lv X
	&=	\mt L\Fn{\bs u_n\fn{\s_n}}\lv x,&
\lv{U_m}\fn{\s_m}
	&=	\mt L\Fn{\bs u_n\fn{\s_n}}\lv{u_m}\fn{\s_m}.
	\label{instantaneous rest frame}
}
In particular, $O_n$ itself is at rest in this frame: $\lv{U_n}\fn{\s_n} = (1, \bs 0)$ at $\lv{X_n}\fn{\s_n}$.

Furthermore, the infinitesimal time lapse in the rest frame coincides with the proper time interval:
\al{
\df\s_n 
	&=	\df X_n^0\sqrt{1-\pn{\df\bs X_n\ov\df X_n^0}^2}
	=	\df X_n^0.
	\label{ds is dX0}
}
Thus, the Lorentz-invariant proper time $\s_n$ matches the time flow experienced by $O_n$.

Since $\alpha_n^0 = 0$ whenever $\bs u_n = 0$ (as shown in Sec.~\ref{acceleration section}), the rest-frame acceleration satisfies
\al{
A_n^0\fn{\s_n} = 0,
}
where
\al{
\lv{A_n}\fn{\s_n}
	&:=	\mt L\Fn{\bs u_n\fn{\s_n}} \lv{\alpha_n}\fn{\s_n}
}
is the covariant acceleration transformed to the instantaneous rest frame.

\subsection{Equation of motion}\label{eom section}
Let $m_n$ be the mass of $O_n$.
In the instantaneous rest frame~\eqref{instantaneous rest frame}, the object~$O_n$ is at rest and hence its motion is governed by a non-relativistic equation of motion:
\al{
m_nc^2{\df\bs U_n\fn{\s_n}\ov\df X_n^0}
	&=	\bs F_n\fn{\s_n},
		\label{non-relativistic}
}
where $\bs F_n\fn{\s_n}$ is the sum of all the non-relativistic forces felt by $O_n$ at $\s_n$;
recall that $\bs U_n$ is dimensionless and that the time-length $X_n^0$ has the dimension of length, hence the extra factor $c^2$.
Since $\df X_n^0$ here is equal to the Lorentz-\emph{invariant} $\df\s_n $ (recall Eq.~\eqref{ds is dX0}), we may obtain its equation of motion in the world frame by
\al{
m_nc^2{\df\lv{u_n}\fn{\s_n}\ov\df\s_n }
	&=	\lv{f_n}\fn{\s_n},
		\label{covariant eom}
}
where
\al{
\lv{f_n}\fn{\s_n}
	&:=	\mt L\Fn{-\bs u_n\fn{\s_n}}\bmat{0\\ \bs F_n\fn{\s_n}},
}
in which we used the inverse expression~\eqref{L inverse}.
We stress that the time component of the equation of motion does not give independent information and can be totally neglected.\footnote{
Indeed, Eq.~\eqref{covariant eom} leads to
\als{
m_nc^2\bmat{{\bs\alpha_n\cdot\bs u_n\ov u_n^0}\\ \bs\alpha_n}
	&=	\bmat{u_n^0&\bs u_n^\t\\\bs u_n&I+\pn{u_n^0-1}\bh u_n\bh u_n^\t}\bmat{0\\ \bs F_n},
}
where we used Eq.~\eqref{alpha0}. The temporal and spatial components read
\als{
m_nc^2{\bs\alpha_n\cdot\bs u_n\ov u_n^0}
	&=	\bs u_n\cdot\bs F_n,&
m_nc^2\bs\alpha_n
	&=	u_n^0\bs F_{n\parallel}
		+\bs F_{n\perp},
}
where
$\bs F_{n\parallel}
	:=	\pn{\bs F_n\cdot\bh u_n}\bh u_n$ and
$\bs F_{n\perp}
	:=	\bs F_n-\bs F_{n\parallel}$.
Noting that $\bh u_n\cdot\bs F_{n\perp}=0$ and hence $\bs u_n\cdot\bs F_n=\bs u_n\cdot\bs F_{n\parallel}$, we see that the temporal component is nothing but the inner product of the spatial component with $\bs u_n$.
}

In electromagnetism, conversely, we can directly derive $\lv{f_n}$ in fully Lorentz-\emph{covariant} fashion, and then derive the expression of the rest-frame force $\bs F_n$ if necessary.
The derivation of~$\lv{f_n}$ is one of the main subjects in the following.

This covariant equation of motion provides the theoretical foundation for updating particle trajectories in our simulation, as we detail in the following implementation sections.

\section{Drawing the world on PLC}\label{sec:Drawing the world on the past light cone}
This section explains how to simulate what a player sees by reconstructing the world on their PLC, using stored worldline data and Lorentz transformations.

\subsection{Definition and visualization of PLC}

\begin{figure}[tp]
\centering
\includegraphics[width=0.4\textwidth]{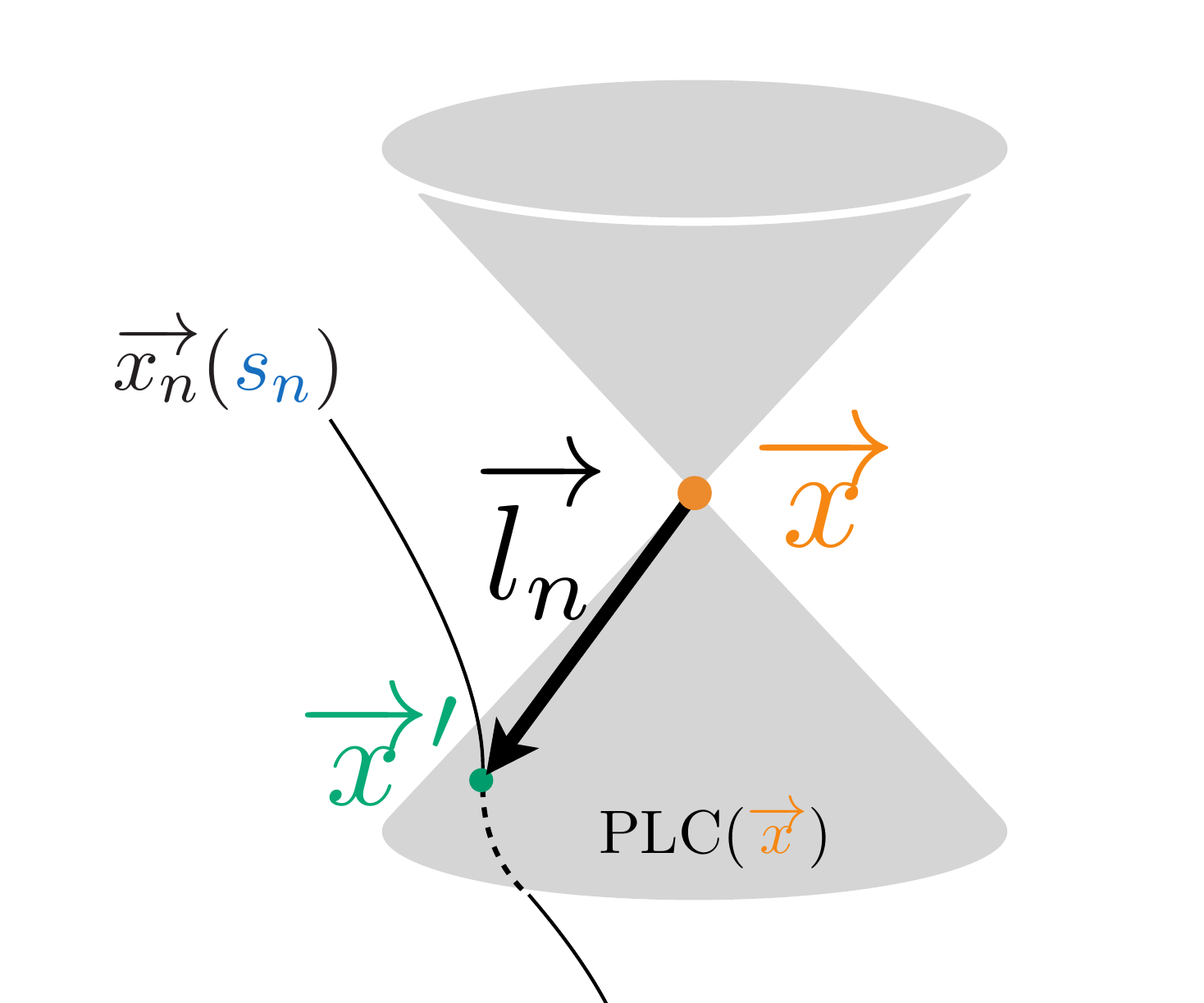}
\caption{
Schematic illustration: 
$\protect\lx$ denotes the spacetime point at which the electromagnetic field is evaluated. 
$\PLC\fn{\protect\lx}$ is its PLC. 
$\protect\lv{x_n}\fn{\sn}$ parametrizes the worldline of the $n$th charge. 
$\green{\protect\lv x'}$ indicates the spacetime location of the charge that influences the field at the observation point $\protect\lx$, and 
$\protect\lv{l_n}$ is the chargeward vector connecting the two points.
}
\label{intersection figure}
\end{figure}

When the player $P$ is at a spacetime point $\lv{x_\P}$ (in the world frame), the world seen by $P$ is the collection of points from which a light ray can reach $\lv{x_\P}$---the PLC:
\al{
\PLC\fn{\lv{x_\P}}
	&:=	\Set{\lv x|\lip{\lv x-\lv{x_\P}}^2=0,\ x^0<x_\P^0}.
}
Note that $\PLC\fn{\lv{x_\P}}$ is defined Lorentz invariantly.\footnote{
The condition $x^0<x^0_\P$ is also invariant under the (orthochronous) Lorentz transformation; see footnote~\ref{proper orthochronous}.
}
An object $O_n$ is observed by the player at the intersection between its worldline $\mc W_n$ and $\PLC\fn{\lv{x_\P}}$; see Fig.~\ref{intersection figure}, where $\lv{x_\P}$ is denoted by $\lx$.
As said above, the player at its proper time length $\s_\P$ sees the world in its instantaneous rest frame:
\al{
\lv X	&=	\mt L\Fn{\bs u_\P\fn{\s_\P}}\lv x.
	\label{player's instantaneous rest frame}
}

\subsection{Intersection with worldlines}\label{worldline section}
\label{worldline section}
Here we explain how to implement the worldline in a computer game.
We let the program store data of the worldlines in the world frame $\lv x$. After $\ttt N$ iterations, $O_n$'s worldline $\mc W_n$ becomes a discrete set of its past spacetime positions:
\al{
	\mathcal{W}_n
	=	\Set{\lv{x_n}_{\pn{\tt i}}}_{\ttt i=\ttt0,\dots,\ttt N}
	=	\Set{\lv{x_n}_{\pn{\tt0}}, \lv{x_n}_{\pn{\tt1}}, \cdots, \lv{x_n}_{\pn{\tt N}}},
	\label{discrete worldline}
}
where we order from the past to the future.

Now we illustrate how to obtain an intersection between a particle's worldline $\mathcal{W}_n$ and $\PLC\fn{\lv{x_\P}}$.
First, we check, from the past to the future, whether $\lv{x_n}_{\pn{\tt i}} \in \mathcal{W}_n$ ($\ttt i=\ttt0,\dots,\ttt N$) fits in the past-side of $\PLC(\lv{x_\P})$:
\al{
	\lip{\lv{x_n}_{\pn{\ttt i}} - \lv{x_\P}}^2 < 0 \quad\text{and}\quad {x_n^0}_{\pn{\tt i}} < x_\P^0.
}
Let $\ttt j$ ($\in\Set{\ttt0,\dots,\ttt N}$) be the first iteration that violates this check. That is, $\lv{x_n}_{\pn{\ttt j-1}}$ and $\lv{x_n}_{\pn{\ttt j}}$ are the last and first spacetime points inside and outside of $\PLC\fn{\lv{x_\P}}$, respectively. 
By linear-interpolating a point $\lv x$ in between $\lv{x_n}_{\pn{\ttt j-1}}$ and $\lv{x_n}_{\pn{\ttt j}}$ as
\al{
\lv x = \pn{1 - \lambda}\lv{x_n}_{\pn{\ttt j-1}} + \lambda \lv{x_n}_{\pn{\ttt j}},\quad  \pn{0\leq \lambda \leq 1},
}
the value of $\lambda$ that gives the intersection point can be determined by the condition $\lip{\lv x - \lv{x_\P}}^2 = 0$:
\al{
\lambda
	&=	{B-\sqrt{B^2-AC}\ov A},
}
where
\al{
A	&:=	-\lip{\lv{x_n}_{\pn{\ttt j}}-\lv{x_n}_{\pn{\ttt j-1}}}^2>0,\\
B	&:=	-\lip{\lv{x_n}_{\pn{\ttt j}}-\lv{x_n}_{\pn{\ttt j-1}},\lv{x_\P}-\lv{x_n}_{\pn{\ttt j-1}}}>0,\\
C	&:=	-\lip{\lv{x_\P}-\lv{x_n}_{\pn{\ttt j-1}}}^2>0.
}
We have discarded the intersection with the future light cone $\pn{B+\sqrt{B^2-AC}}/A$.

If necessary, the $O_n$'s velocity on $\PLC\fn{\lv{x_\P}}$, which is nothing but the velocity from $\lv{x_n}_{\pn{\ttt j-1}}$ to $\lv{x_n}_{\pn{\ttt j}}$, can be computed as
\al{
	\lv{u_n}_{\pn{\ttt j-1}} &:= {\lv{\Delta x_n}_{\pn{\ttt j-1}}\ov \Delta\s_{n\pn{\ttt j-1}}},
}
where
\al{
\lv{\Delta x_n}_{\pn{\ttt j-1}}
	&:= \lv{x_n}_{\pn{\ttt j}} - \lv{x_n}_{\pn{\ttt j-1}},\\
\Delta\s_{n\pn{\ttt j-1}}
	&:=	\sqrt{-\lip{\lv{\Delta x_n}_{\pn{\ttt j-1}}}^2}=\sqrt{-\lip{\lv{x_n}_{\pn{\ttt j}} - \lv{x_n}_{\pn{\ttt j-1}}}^2}.
}
Accordingly, one may also store the information of the proper time length of $O_n$ given by
\al{
{\s_n}_{\pn{\ttt j}}
	&:=	{\s_n}_{\pn{\ttt j-1}}+\Delta\s_{n\pn{\ttt j-1}}
}
in $\mc W_n$.

Given $\lv{x_n}_{\pn{\ttt j}}$, $\lv{x_n}_{\pn{\ttt j-1}}$, and $\lv{x_n}_{\pn{\ttt j-2}}$, the covariant acceleration can be approximated by
\al{
\lv{\alpha_n}_{\pn{\ttt j-1}}
	&:=	{\lv{u_n}_{\pn{\ttt j-1}}-\lv{u_n}_{\pn{\ttt j-2}}\ov {\Delta\s_{n\pn{\ttt j-1}}+\Delta\s_{n\pn{\ttt j-2}}\ov2}}\nn
	&=	{2\ov\Delta\s_{n\pn{\ttt j-1}}+\Delta\s_{n\pn{\ttt j-2}}}
		\pn{
			{\lv{x_n}_{\pn{\ttt j}}-\lv{x_n}_{\pn{\ttt j-1}}\ov \Delta\s_{n\pn{\ttt j-1}}}
			-{\lv{x_n}_{\pn{\ttt j-1}}-\lv{x_n}_{\pn{\ttt j-2}}\ov \Delta\s_{n\pn{\ttt j-2}}}
			}.
}

\subsection{Drawing the observed world}
Once the intersecting points $\Set{\lv{x_n}}_{n=1,\dots,N}$ with $\PLC\fn{\lv{x_\P}}$ are obtained for $N$ point charges, they are transformed to the player's instantaneous rest frame. As in Eq.~\eqref{player's instantaneous rest frame}, this transformation is given by
\al{
\lv{X_n}
	&=	\mt L\fn{\bs u_\P} \pn{\lv{x_n} - \lv{x_\P}},
		\label{transformation to rest frame}
}
where $\bs u_\P$ is the player's (dimensionless covariant) velocity at $\lv{x_\P}$ in the world frame, and we have subtracted $\lv{x_\P}$ to place the player at the origin of the new coordinate system.

\subsection{Time evolution of player}\label{time evolution section}
We present a schematic time evolution for the player to account for where $c$ appears.

In our implementation, the real-world time is synchronized with the player's proper time $\tau_\P$, or the corresponding proper time length $s_\P=c\,\tau_\P$.
At the real-world time $t$, we have $s_\P=ct$.
Let $\Delta t$ be the real-world time-lapse in the next iteration.
Then the player's proper-time-length lapse is $\Delta s_\P=c\Delta t$.
In the next iteration, the player's position and velocities change into, in the world frame,\footnote{
The time component of the velocity can be fixed as $u^0_\P\fn{s_\P+\Delta s_\P}=\sqrt{1+\bs u_\P^2\fn{s_\P+\Delta s_\P}}$.
}
\al{
\lv{x_\P}\fn{s_\P+\Delta s_\P}
	&=	\lv{x_\P}\fn{s_\P}+\lv{u_\P}\fn{s_\P}\Delta s_\P,
		\label{x evolution}\\
\bs u_\P\fn{s_\P+\Delta s_\P}
	&=	\bs u_\P\fn{s_\P}+\bs\alpha_\P\fn{s_\P}\Delta s_\P,
		\label{u evolution}
}
where the acceleration is given as
\al{
\lv{\alpha_\P}\fn{s_\P}
	&=	{\lv{f_\P}\fn{s_\P}\ov m_\P c^2}
		\label{player's acceleration}
}
from the equation of motion~\eqref{covariant eom}.\footnote{
If the player is charged, the player will feel the electromagnetic force~\eqref{eom simple form} below, which should also be added in the force in Eq.~\eqref{player's acceleration}.
}

From the dimesionalities
\al{
\sqbr{\Delta s_\P}
	&=	\sqbr{c}\sqbr{\tx{Time}},&
	\bigl(&=	\sqbr{\tx{Length}}\,\bigr)\\
\sqbr{\bs u_\P}
	=	\sqbr{\lv{u_\P}}
	&=	\sqbr{\tx{Non-relativistic velocity}\ov c},&
	\bigl(&=	\sqbr{\tx{Dimensionless}}\,\bigr)\\
\sqbr{\bs\alpha_\P}
	=	\sqbr{\lv{\alpha_\P}}
	&=	\sqbr{\tx{Non-relativistic acceleration}\ov c^2},&
		\bigl(&=	\sqbr{1\ov\tx{Length}}\,\bigr)
}
we see that the time evolution in Eqs.~\eqref{x evolution} and \eqref{u evolution} has the same dimensionality as the ordinary non-relativistic one.

In actual implementation, the Euler method in Eqs.~\eqref{x evolution} and \eqref{u evolution} is known to increase the total energy of the system exponentially. Consequently, this method results in pathological behaviors such as perpetually rotating opposite charges despite the emission of electromagnetic waves. This issue can be mitigated by employing the symplectic (semi-implicit) Euler method:
\al{
\bs u_\P\fn{s_\P+\Delta s_\P}
	&=	\bs u_\P\fn{s_\P}+\bs\alpha_\P\fn{s_\P}\Delta s_\P,\\
\lv{x_\P}\fn{s_\P+\Delta s_\P}
	&=	\lv{x_\P}\fn{s_\P}+\lv{u_\P}\fn{s_\P+\Delta s_\P}\Delta s_\P.
}
We utilize this symplectic Euler method in the sample code explained in Sec.~\ref{Implementation}.

\subsection{Time evolution of others}\label{time evolution section}

At $s_\P$, the program draws the world on $\PLC\fn{\lv{x_\P}\fn{s_\P}}$.
In the next iteration, the program reads out the real-world time-lapse, identified as $\Delta\tau_\P$, or $\Delta s_\P=c\,\Delta\tau_\P$.
Then, we need to time-evolve all the point charges up to $\PLC\fn{\lv{x_\P}\fn{s_\P+\Delta s_\P}}$.
For a consistent time evolution in accordance with causality, we employ the following algorithm, which is newly explained here but has already been implemented in Ref.~\cite{10.1093/ptep/ptx127}.

Let $\Set{\lv{x_n}\fn{s_n}}_{n=1,\dots,N}$ be the last intersecting points of the $N$ worldlines with $\PLC\fn{\lv{x_\P}\fn{s_\P}}$; see Fig.~\ref{intersection figure}.
We want to evolve the $N$ charges consistently with causality. 
We first move a past-most charge with a pre-fixed small time step $\Delta x^0$ in the world frame.\footnote{
In the implementation of Ref.~\cite{10.1093/ptep/ptx127}, the interplay among the player and non-player characters was important. Therefore, a fixed proper-time length $\Delta s_n$ was used. In the current implementation, a charge obeys only the Lorentz force and does not change its move by some extra acceleration of its own. So it suffices to consider fixed world-frame time lapse.
}
We then evolve the past-most one after each iteration until all the charged particles reach $\PLC\fn{\lv{x_\P}\fn{s_\P+\Delta s_\P}}$.

At each of the iterations, the position and velocity of the charge are determined by the Lorentz force described below.

\section{Relativistic electromagnetism}\label{elemag review section}
We briefly review the relativistic electromagnetism.
Once again, we ensure to include details that, although might be elementary for experts in physics and mathematics, are crucial for making the content accessible to a broader audience.

\subsection{Equation of motion}
In the electromagnetic dynamics, we promote the equation of motion~\eqref{non-relativistic} (of $O_n$ at a spacetime point $\lv{x_n}$ with velocity $\bs u_n$) to\footnote{
In a more familiar-looking expression,
$
c{\df\ov\df x_n^0}={\df\ov\df t}
$.
In the non-relativistic regime, $c\bs u_n\approx\bs v_n$, Eq.~\eqref{eom of O} reduces to the non-retivistic equation of motion $m_n{\df\bs v_n\ov\df t}\approx\bs f_n\fn{\lv{x_n}}$.
Here, we write the left-hand side of the equation of motion~\eqref{eom of O} in the form that is correct in the relativistic theory from the beginning.
On the other hand, the Lorentz force~\eqref{Lorentz force} does not need such a modification to be relativistic.
}
\al{
m_nc^2{\df\bs u_n\ov\df x_n^0}
	&=	\bs f_n^\tx{Lorentz}\fn{\lv{x_n}},
		\label{eom of O}
}
where $\bs f_n^\tx{Lorentz}$ is the Lorentz force felt by $O_n$:
\al{
\bs f_n^\tx{Lorentz}\fn{\lv{x_n}}
	&:=	q_n\sqbr{\bs E\fn{\lv{x_n}}+\bs v_n\times\bs B\fn{\lv{x_n}}},
		\label{Lorentz force}
}
in which $q_n$ is the charge of $O_n$;
$\bs E$ and $\bs B$ are the electric and magnetic fields at the location of the charge, respectively;
$\bs v_n$ is the non-covariant velocity~\eqref{non-covariant velocity given}; and
the vector product is defined by, for $i=1,2,3$,
\al{
\pn{\bs A\times\bs B}_i
	&=	\sum_{j,k=1}^3\epsilon_{ijk}A_jB_k,&
\bmat{
\pn{\bs A\times\bs B}_1\\
\pn{\bs A\times\bs B}_2\\
\pn{\bs A\times\bs B}_3
}
	&=	\bmat{
			A_2B_3-A_3B_2\\
			A_3B_1-A_1B_3\\
			A_1B_2-A_2B_1\\
			},
}
where
\al{
\epsilon_{ijk}
	&=	\begin{cases}
		1	&	\tx{(when $i,j,k$ is an even permutation of $1,2,3$)},\\
		-1	&	\tx{(when $i,j,k$ is an odd permutation of $1,2,3$)},\\
		0	&	\tx{(otherwise)},
		\end{cases}
}
is the Levi-Civita symbol, namely, the totally anti-symmetric tensor for the (spatial) rotational group $SO(3)$.
With our metric covention~\eqref{eta defined}, the upper and lower spatial indices are not distinguished:
\al{
A^i
	&=	A_i,&
\epsilon^{ijk}
	&=	\epsilon_{ijk},
	\label{irrelevance of upper and lower positions}
}
etc.; see Eq.~\eqref{raising snd lowering indices} below.

It is crucial to note that once the charge $q_n$ is specified, the motion of $O_n$ is governed by the electromagnetic fields. These fields are determined at any spacetime position of the charge, denoted by $\lv{x_n}$.

Given the motion of the objects $\Set{O_n}_{n=1,2,\dots}$, namely their world lines $\Set{\mc W_n}_{n=1,2,\dots}$, the charge and (3D) current densities $\rho\fn{\lx}$ and $\bs j\fn{\lx}$ at a spacetime point $\lx$ are obtained as
\al{
\rho\fn{\lx}
	&=	\sum_nq_n\int\df\sn \,\delta^4\Fn{\lx-\lv{x_n}\fn{\sn }},\\
\bs j\fn{\lx}
	&=	\sum_nq_n\int\df\sn \,\delta^4\Fn{\lx-\lv{x_n}\fn{\sn }} c\bs u_n\fn{\sn }.
}
Recall that $\bs u_n\fn{\sn}$ is dimensionless, and hence the extra factor $c$.
Here and hereafter, we sometimes put colors for ease of eyes.

These charge and current densities determine the electromagnetic fields via Maxwell's equations at each spacetime point $\lx$:
\al{
\nab\cdot\bs E\fn{\lx}
	&=	{\rho\fn{\lx}\ov\epsilon_0},&
\nab\times\bs B\fn{\lx}
	&=	{\bs j\fn{\lx}\ov\epsilon_0c^2}
		+{1\ov c^2}c\p_0\bs E\fn{\lx},
		\label{matter interaction}\\
\nab\cdot\bs B\fn{\lx}
	&=	0,&
\nab\times\bs E\fn{\lx}
	&=	-c\p_0\bs B\fn{\lx}.\label{purely electromagnetic}
}
where, for $\mu=0,\dots,3$,
\al{
\p_\mu
	&:=	{\p\ov\p x^\mu},&
c\p_0
	&=	c{\p\ov\p x^0}={\p\ov\p t},
}
and $\epsilon_0$ is the electric constant (vacuum permittivity).

The physical insight is most simply obtained by taking the natural units
\al{
\epsilon_0=c=1,
}
which also yields the natural unit for the magnetic constant (vacuum permeability)
$\mu_0:={1\ov\epsilon_0c^2}=1$.
Although these constants can be easily recovered when necessary by dimensional analysis,
we take the non-natural units and leave them as they are for readers unfamiliar with the dimensional analysis.
Finally, even if one varies $c$, as in our sample program that will be described in Sec.~\ref{Implementation}, one can still safely take
\al{
\epsilon_0=1
	\label{epsilon0=1 unit}
}
(yielding $\mu_0=1/c^2$), unless one further varies $\epsilon_0$.

Note that the tensor $\p_\mu$ transforms under the inverse representation of the Lorentz group. That is, when the coordinates transforms according to Eq.~\eqref{Lorentz transformation on x}, we obtain
\al{
\p_\mu
={\p\ov\p x^\mu}
	&\to	\p'_\nu
	=	{\p\ov\p x^{\pr\mu}}
	=	\sum_{\nu=1}^3{\p x^\nu\ov\p x^{\pr\mu}}{\p\ov\p x^\nu}
	=	\sum_{\nu=1}^3\bmat{\mt\Lambda^{-1}}\!{}^\nu{}_\mu\p_\nu,
}
where we used
\al{
{\p x^{\pr\mu}\ov\p x^\nu}
	&=	\Lambda^\mu{}_\nu,&
{\p x^\nu\ov\p x^{\pr\mu}}
	&=	\bmat{\mt\Lambda^{-1}}\!{}^\nu{}_\mu.
	\label{derivative inverse}
}
We always write an inverse representation of the Lorentz group, such as $\p_\mu$, with the lower indices so that the contraction of all the lower and upper indices gives a Lorentz invariant.

The purely electromagnetic part~\eqref{purely electromagnetic} is automatically solved when we rewrite the magnetic and electric fields in terms of the scalar and (3D) vector potentials $\phi\fn{\lx}$ and $\bs A\fn{\lx}$:
\al{
\bs B\fn{\lx}
	&=	\nab\times\bs A\fn{\lx},&
\bs E\fn{\lx}
	&=	-\nab\phi\fn{\lx}-c\p_0\bs A\fn{\lx}.
		\label{BE in terms of Aphi}
}
From $\pn{\phi,\bs A}$ and $\pn{\rho,\bs j}$, we respectively define a (covariant) vector potential $\lv A$, which we call the \emph{gauge field} hereafter, and a (covariant) current density $\lv j$:
\al{
\lv A\fn{\lx}
	&=	\Pn{A^\mu\fn{\lx}}_{\mu=0,\dots,3}
	:=	\pn{{\phi\fn{\lx}\ov c},\bs A\fn{\lx}},\\
\lv j\fn{\lx}
	&=	\Pn{j^\mu\fn{\lx}}_{\mu=0,\dots,3}
	:=	\Pn{c\rho\fn{\lx},\bs j\fn{\lx}}.
}
In the relativistic electrodynamics, we assume that they transform covariantly under the Lorentz transformation.

For any tensor such as $A^\mu$ and $\p_\mu$, we define
\al{
A_\mu
	&:=	\sum_{\nu=0}^3\eta_{\mu\nu}A^{\nu},&
\p^\mu
	&:=	\sum_{\nu=0}^3\eta^{\mu\nu}\p_{\nu},
	\label{raising snd lowering indices}
}
etc. That is, $A_0:=-A^0$, $\p^0:=-\p_0$ and $A_i:=A^i$, $\p^i=\p_i$ ($i=1,2,3$). Since $\eta$ is the invariant tensor of the Lorentz group in the sense of Eq.~\eqref{Lambda as Lorentz transformation}, we see that $A_\mu$ transforms under the inverse representation of the Lorentz group, whereas $\p^\mu$ the same as $x^\mu$.
In this notation, the inverse transformation~\eqref{Lambda inverse explicitly written}, or \eqref{derivative inverse}, reads $\bmat{\mt\Lambda^{-1}}\!{}^\nu{}_\mu=\Lambda_\mu{}^\nu$.

We define a rank-two anti-symmetric tensor, the \emph{field strength},
\al{
F_{\mu\nu}\fn{\lx}
	&:=	\p_\mu A_\nu\fn{\lx}-\p_\nu A_\mu\fn{\lx}.
		\label{field strength defined}
}
We see that the electromagnetic fields can be written in terms of the field strength: For $i=1,2,3$, Eq.~\eqref{BE in terms of Aphi} reads
\al{
E_i\fn{\lx}
	&=	cF_{i0}\fn{\lx}
	=	cF^{0i}\fn{\lx},\nn
B_i\fn{\lx}
	&=	{1\ov2}\sum_{j,k=1}^3\epsilon_{ijk}F_{jk}\fn{\lx}
	=	{1\ov2}\sum_{j,k=1}^3\epsilon_{ijk}F^{jk}\fn{\lx}.
		\label{electromagnetic field}
}
More explicitly, the magnetic reads
\al{
B_1\fn{\lx}
	&=	F_{23}\fn{\lx},&
B_2\fn{\lx}
	&=	F_{31}\fn{\lx},&
B_3\fn{\lx}
	&=	F_{12}\fn{\lx}.	
	\label{B from F more concrete}
}
Now the equation of motion under the Lorentz force~\eqref{eom of O} can be written in a manifestly Lorentz-covariant (and invariant) fashion: For $\mu=0,\dots,3$ and for each $n$,
\al{
m_n c{\df u_n^\mu\fn{\sn}\ov\df\sn }
	&=	q_n\sum_{\rho,\sigma=0}^3F^{\mu\rho}\Fn{\lv{x_n}\fn{\sn}}\eta_{\rho\sigma}u^\sigma\fn{\sn},
}
or more concisely,
\al{
m_n c{\df u_n^\mu\fn{\sn}\ov\df\sn }
	&=	q_n\sum_{\nu=0}^3F^{\mu\nu}\Fn{\lv{x_n}\fn{\sn}}u_\nu\fn{\sn},
}
where we have recovered the dependence on the proper time length that was implicit in Eq.~\eqref{eom of O}.\footnote{
We have used ${\df\ov\df x_n^0}={1\ov u_n^0}{\df\ov\df s_n}$ in Eq.~\eqref{eom of O}:
\als{
m_nc^2\ub{\df u_n^i\ov\df x_n^0}_{{1\ov u_n^0}{\df u_n^i\ov\df s_n}}
	=	q_n\Bigg[
			cF^{0i}+\sum_{j,k=1}^3\epsilon_{ijk}\ub{c{\df x_n^j\ov\df x_n^0}}_{c{u_n^j\ov u_n^0}}\pn{\sum_{l,m=1}^3{1\ov2}\epsilon_{klm}F_{lm}}
			\Bigg].
}
}

In the matrix notation (see Eqs.~\eqref{matrix F} and \eqref{transformation of F} below), the above equation of motion reads
\al{
m_n c{\df\lv{u_n}\fn{\sn}\ov\df\sn }
	&=	q_n\mt F\Fn{\lv{x_n}\fn{\sn}}\,\mt\eta\,\lv u\fn{\sn}.
		\label{eom simple form}
}
This form may be more usable in practical implementation.

\subsection{General solution to relativistic Maxwell's equation}

The general form of the gauge field is obtained as follows~\cite{Nakayama:2024wfi}:
\al{
	A^\mu\fn{\lx}
	&= {1\ov 4\pi \epsilon_0 c^2}\int\df^4\green{\lv x'} 
		\delta\fn{\xz-\green{x^{\pr0}} -  \ab{\bx - \green{\bs{x^\pr}}}}
	{j^\mu\fn{\green{\lv x'}}\ov\ab{\bx - \green{\bs{x^\pr}}}}.
		\label{A concrete}
}
Hereafter, we sometimes write the integral more concisely
\al{
\int_{\PLC\fn{\lx}} \df^3\green{\bs{x'}}\sqbr{\cdots}
	\,:=\,
		\int\df^4\green{\lv x'} 
		\delta\fn{\xz-\green{x^{\pr0}} -  \ab{\bx - \green{\bs{x^\pr}}}}
		\sqbr{\cdots}
}
such that
\al{
A^\mu\fn{\lx}
	&=	{1\ov 4\pi\epsilon_0c^2}\int_{\PLC\fn{\lx}} \df^3\green{\bs{x'}} {j^\mu\fn{\green{\lv x'}} \ov \left|\bx - \green{\bs x'}\right|}.
	\label{general solution of A}
}
Physically, $\green{\bs x'}$ is each location of the charge in the past that affects the electromagnetic field at $\lx$ in the future; see Fig.~\ref{intersection figure}.
Here, the restriction of $\green{\lv x'}$ onto $\PLC\fn{\lx}$ is achieved by imposing the time difference $\xz-\green{x^{\pr0}}$ to be equal to the distance $\ab{\bx-\green{\bs x'}}\geq0$, by the delta function.

\section{Covariant formalism for field strength from point charges}\label{covariant formalism section}
We compute how the relativistic motion of a charged particle affects the electromagnetic field in the future. In other words, we compute a fully relativistic expression of the Li\'enard-Wiechert potential in terms of covariant quantities only.
Our main goal is to find out the field strength at $\orange\lx$ in terms of the positions and velocities of point charges on $\PLC\fn{\orange\lx}$.
A reader who is interested only in its final form may skip to Eqs.~\eqref{F0i} and \eqref{Fij}.

\subsection{Continuous worldlines and covariant current density}
The electromagnetic field at a spacetime point $\lx$ is determined by summing over the influence from each object $O_n$ on $\PLC\fn{\lx}$, namely at the intersection between $\mc W_n$ and $\PLC\fn{\lx}$. Each influence is solely determined by the position $\lv{x_n}$ and velocity $\lv{u_n}$ on $\PLC\fn{\lx}$.
In the actual implementation on computers, we may always obtain them from the discrete worldline as in Sec.~\ref{worldline section}. Therefore, we take $\lv{x_n}$ and $\lv{u_n}$ for granted, and compute its electromagnetic influence on $\lx$.
This way, we treat each worldline~\eqref{discrete worldline} as if it were continuously parametrized by its proper time length $\sn $:
\al{
\mc W_n
	&=	\Set{\lv{x_n}\fn{\sn }\mid-\infty<\sn <\infty}.
}

Given the worldlines $\Set{\mc W_n}_{n=1,2,\dots}$ of the charged particles, the covariant current density at a spacetime point $\green{\lv x'}$ is given by
\al{
\lv j\fn{\green{\lv x'}}
	&=	\sum_ncq_n\int\df\sn \,\delta^4\fn{\green{\lv x'}-\lv{x_n}\fn{\sn }}\lv{u_n}\fn{\sn },
	\label{current j}
}
where $q_n$ is the charge of $O_n$.
Here, $c$ is supplied to make the dimension of the current density $\pn{\tx{charge}}/\pn{\tx{length}}^2\pn{\tx{time}}$.

\subsection{Modified gamma factor and chargeward vector}

To write down the field strength below, we define a \emph{modified gamma factor}:
\al{
\gamma_n\fn{\sn,\bx}
	&:=	{\df x_n^0\fn{\sn}\ov\df\sn}+{\p\ab{\bx-\bs x_n\fn{\sn}}\ov\p\sn}\nn
	&=	u_n^0\fn{\sn }+\bh l_n\fn{\sn,\bx}\cdot\bs u_n\fn{\sn },
	\label{gamma_n defined}
}
where we define the \emph{chargeward vector} as the covariant position vector of the charge relative to the reference point:
\al{
\lv{l_n}\fn{\sn,\lx}
	&:=	\lv{x_n}\fn{\sn}-\lx,&
&\tx{namely,}&
\bmat{l_n^0\fn{\sn,\xz}\\ \bs l_n\fn{\sn,\bx}}
	&:=	\bmat{x_n^0\fn{\sn}-\xz \\ \bs x_n\fn{\sn }-\bx};
			\label{ln given}
}
see Fig.~\ref{intersection figure}. Recall also the notation~\eqref{hat notation}:
\al{
\bh l_n\fn{\sn,\bx}
	&=	{\bs l_n\fn{\sn,\bx}\ov\ab{\bs l_n\fn{\sn,\bx}}}.
		\label{ln hat given}
}
Note that the modified gamma factor is written in terms of the covariant quantities $\lv{u_n}$ and $\lv{l_n}$.

\subsection{Field strength}\label{time derivative section}

The components of field strength $F_{i0}=-F_{0i}=F^{0i}=-F^{i0}$ and $F_{ij}=-F_{ji}=-F^{ji}=F^{ij}$ are obtained as follows~\cite{Nakayama:2024wfi}:
\al{
F^{0i}\fn{\lx}
	&=	-\p_0A^i\fn{\lx}-\p_iA^0\fn{\lx}\nn
	&=	\sum_n{q_n\ov4\pi\epsilon_0c\ab{\bs l_n}}
		\sqbr{
			\h l^i{
						u_n^0
							\pn{
								\bh l_n\cdot\bs\alpha_n
								-{1\ov\ab{\bs l_n}}
								}
						-\alpha_n^0\pn{\bh l_n\cdot\bs u_n}
					\ov\gamma_n^3
						}
			+u_n^i{
			\alpha_n^0+\bh l_n\cdot\bs\alpha_n
			-{1\ov\ab{\bs l_n}}
			\ov\gamma_n^3
			}
			-{\alpha_n^i\ov\gamma_n^2}
		},\label{F0i}\\
F^{ij}\fn{\lx}
	&=	\p_iA^j\fn{\lx}-\p_jA^i\fn{\lx}\nn
	&=	\sum_n{q_n\ov4\pi\epsilon_0c\ab{\bs l_n}}\sqbr{
			{\h l_n^i\alpha_n^j-\h l_n^j\alpha_n^i\ov\gamma_n^2}
			-{\h l_n^iu_n^j-\h l_n^ju_n^i\ov\gamma_n^3}\pn{\alpha_n^0+\bh l_n\cdot\bs\alpha_n-{1\ov\ab{\bs l_n}}}
			},
			\label{Fij}
}
where $\bs l_n$ and $\bh l_n$ are given in Eqs.~\eqref{ln given} and \eqref{ln hat given}, respectively, and $\lv{u_n}$, $\lv{\alpha_n}$, and $\gamma_n$ are all evaluated at the intersection between $\mc W_n$ and $\PLC\fn{\lx}$.
This expression for the field strength can be used directly in computer programs to obtain the Lorentz transformation of the electromagnetic fields below.

Given the field strength, the electric and magnetic fields can be derived from Eq.~\eqref{electromagnetic field}.
For an actual implementation in a computer program, one may write the field strength as an anti-symmetric matrix $\mt F$ whose $\mu,\nu$ components are given by the upper-indexed counterparts:
\al{
\bmat{\mt F\fn{\lx}}^{\mu,\nu}
	&=	F^{\mu\nu}\fn{\lx}.
	\label{matrix F}
}
Then its Lorentz transformation law under the coordinate transformation $\lx\to\orange{\lx'}=\mt\Lambda\lx$ is
\al{
\mt F
	&\to	\mt F'=\mt\Lambda\mt F\mt\Lambda^\t.
	\label{transformation of F}
}
Accordingly, the Lorentz transformation for the electromagnetic fields are
\al{
E^i
	&\to	E^{\pr i}
	=	\red cF^{\pr0i}
	=	\red c\bmat{\mt\Lambda\mt F\mt\Lambda^\t}^{0,i},\\
B^i
	&\to	B^{\pr i}
	=	{1\ov2}\sum_{j,k=1}^3\epsilon_{ijk}F^{\pr jk}
	=	{1\ov2}\sum_{j,k=1}^3\epsilon_{ijk}\bmat{\mt\Lambda\mt F\mt\Lambda^\t}^{j,k};
}
recall Eqs.~\eqref{irrelevance of upper and lower positions} and \eqref{B from F more concrete}.

\section{Overview of concrete implementation}\label{Implementation}
We have developed a sample code to demonstrate a practical implementation~\cite{github}. Below, we describe its functionality.

In our visualization of electric and magnetic fields, they are represented by green and yellow arrows, respectively. 
The electromagnetic fields are evaluated at lattice points, initially located on two planes perpendicular to each other, on the player's PLC.
The speed of light $c$ is expressed in units of this lattice grid as $\tx{grid}/\tx s$.

Positive and negative charges are represented by red and blue regular icosahedrons, respectively. We may define an arbitrary unit charge, say the elementary charge $e$, to be dimensionless: In the sample program, we choose to set $e/4\pi\epsilon_0=1$ for simplicity of the code, while retaining the dimensionality of $c$.\footnote{
The fine structure constant is dimensionless and is represented as $\alpha=e^2/4\pi\epsilon_0\hbar c\simeq1/137.$, where $e$ is the elementary charge and $\hbar c\simeq 0.20\,\tx{GeV}\,\tx{fm}$ (in SI units, $e\simeq1.6\times10^{-19}\,\tx{C}$ and $\hbar c\simeq3.2\times10^{-26}\,\tx{J}\,\tx{m}$).
Therefore, in natural units $\hbar=c=\epsilon_0=1$, the elementary charge $e$ becomes dimensionless and has a fixed value $e\simeq0 .30$.
On the other hand, in the sample program, we take $\epsilon_0=1$ (see Eq.~\eqref{epsilon0=1 unit}) and leave $c$ dimensionful, while setting $e=4\pi\epsilon_0$ ($=4\pi$).
}
In the sample program, we choose $q_n$ to be $q_n=\pm e$ for the positive and negative charges; the values of the charges can be modified in the code if necessary.

\subsection{Speed of Light}
We have implemented a slider that adjusts the value of $c$ by factors of $2^\textsf n$, where the integer $\textsf n$ ranges from $-3$ to 10. When $c$ is increased, the simulation continues unchanged, while when $c$ is decreased, the simulation is reset to its initial configuration so that the speed of massive charges never exceed $c$.

We have also implemented a button that allows users to disable the Lorentz transformation to the player's rest frame~\eqref{transformation to rest frame}. Note that even with the Lorentz transformation disabled, the world is still viewed from the player's PLC.

\subsection{Arrow}
The length \(\ell\) of each arrow is defined such that the electric field at a distance of 1 grid unit from the point charge is represented by an arrow with length 1 grid unit when the Log Reduction Count \(N=0\) and the 10-Exponent \(\mathsfit{n}=0\), both of which are explained below. The length of the arrows for the magnetic field is defined to match that for the electric field when \(c=1\,\tx{grid}/\tx{s}\) and when \(\ab{\bs{B}}=\ab{\bs{E}}/\tx{c}\), such as when \(\bs{E}\) is perpendicular to the line of sight \(\bh{l}\).

We also provide an option for displaying the Poynting vector
\al{
\bs S
	&:=	\epsilon_0c^2\bs E\times\bs B
}
with magenta arrows; see also Ref.~\cite{Nakayama:2024wfi}. The length of the arrows for the Poynting vector is defined such that it becomes the same as that for the electric field when \(c=1\,\tx{grid}/\tx{s}\) and when \(\bs{E}\) and \(\bs{B}\) are perpendicular to each other, namely when \(\ab{\bs{S}}=\epsilon_0c^2\ab{\bs{B}}\ab{\bs{E}}\).

Given that \(\ell\) can vary significantly across different lattice points, we have implemented a ``Log Reduction'' scaling option to manage this variability. This scaling applies the transformation
\al{
\ell \mapsto \ln\fn{1+\ell}
}
iteratively, \(N\) times, where \(\ln\) represents the natural logarithm. We refer to \(N\) as the ``Log Reduction Count.'' Additionally, we implement scaling by an extra factor of \(10^n\) before applying the Log Reduction, where \(n\) is referred to as the ``10-Exponent.''

\begin{figure}[tp]
\centering
\includegraphics[width=0.49\textwidth]{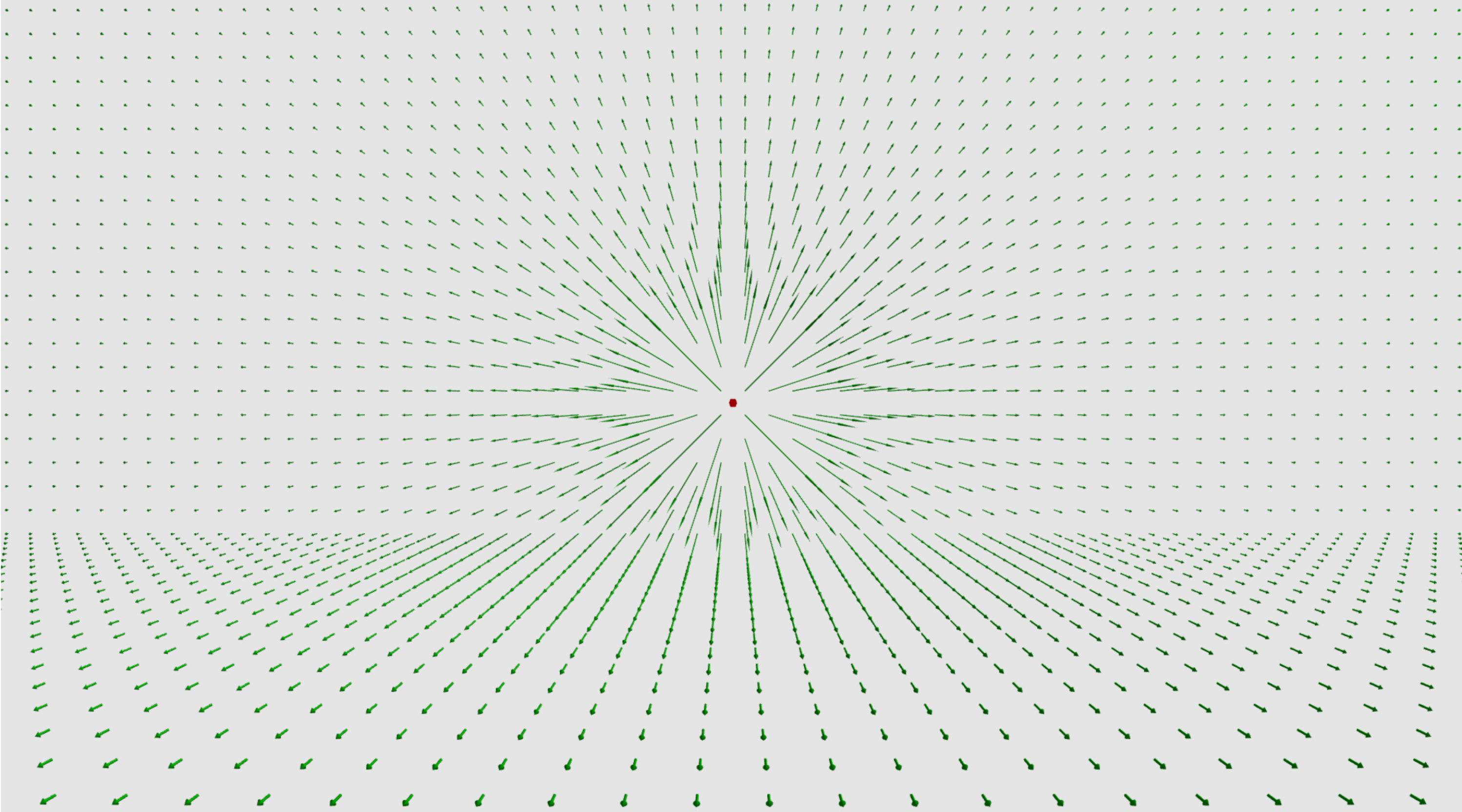}\hfill
\includegraphics[width=0.49\textwidth]{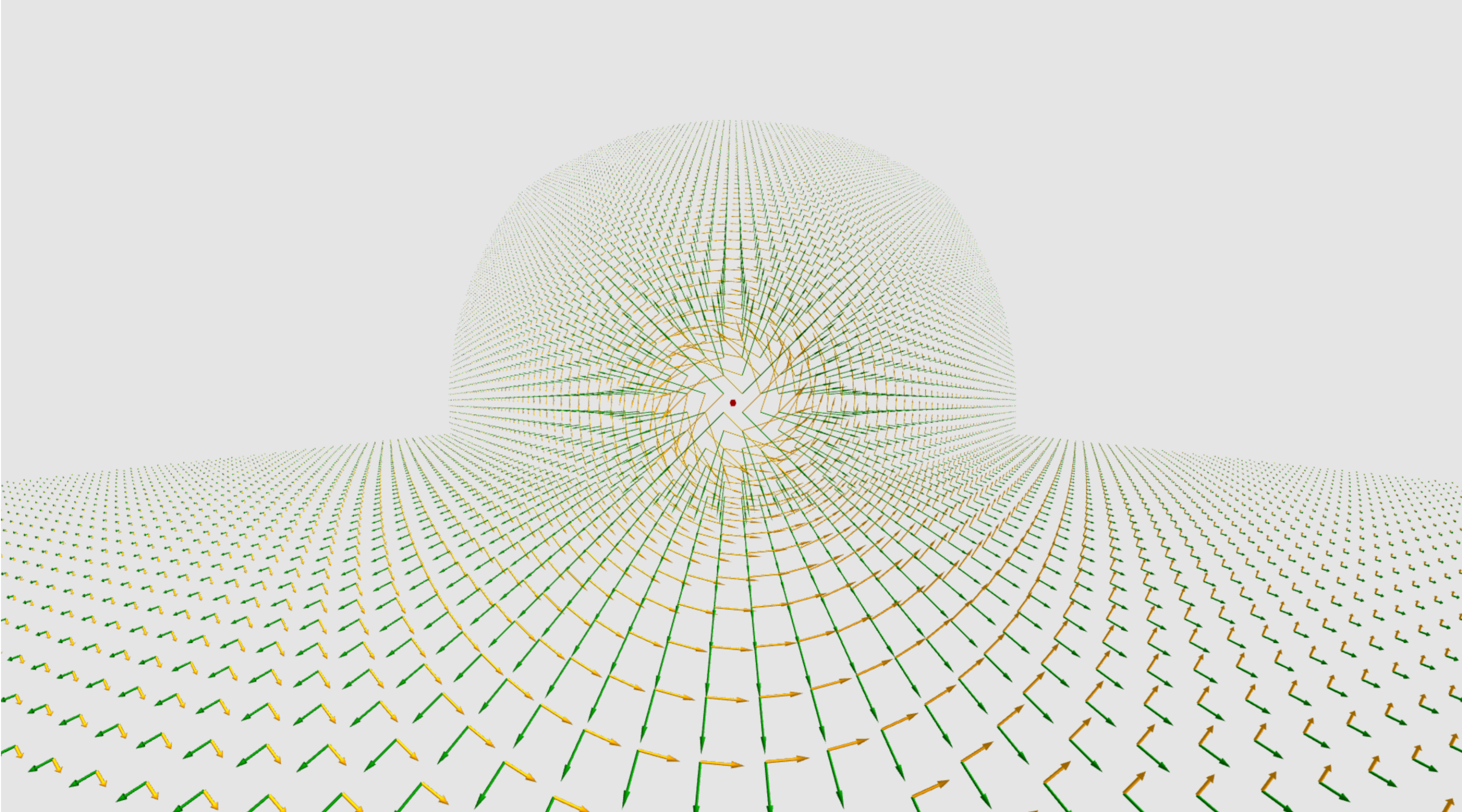}\mbox{}
\caption{
Scene of the sample program with the Static Charge preset. Electromagnetic fields are depicted on vertical and horizontal planes. The setup is $c=1\,\tx{grid}/\tx{s}$, Log Reduction Count \(N=1\), and 10-Exponent $n=2$.
\textbf{Left:} Static Coulomb electric fields (green arrows) emanate from the positive point charge at rest (red icosahedron). 
\textbf{Right:} After forward acceleration of the player, the point charge appears to move toward the player, generating an electric current; this produces magnetic fields (yellow arrows) circulating around the charge's velocity vector toward the player. The planes are Lorentz contracted due to the player acceleration, actually elongated along the direction of acceleration; see Appendix~B in Ref.~\cite{10.1093/ptep/ptx127} for a detailed explanation.
\label{static_charge}}
\end{figure}

\begin{figure}[tp]
\centering
\includegraphics[width=0.49\textwidth]{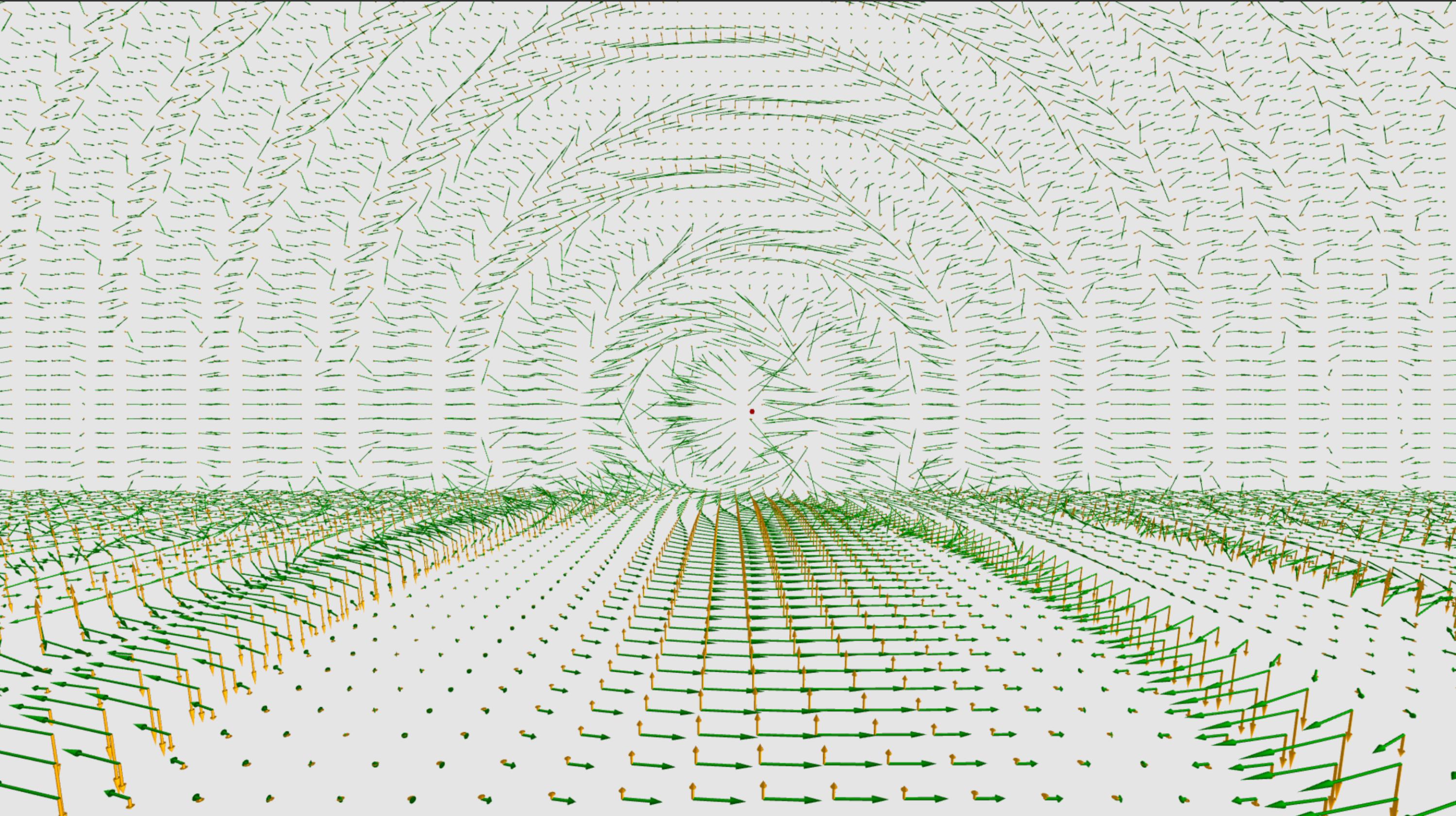}\hfill
\includegraphics[width=0.49\textwidth]{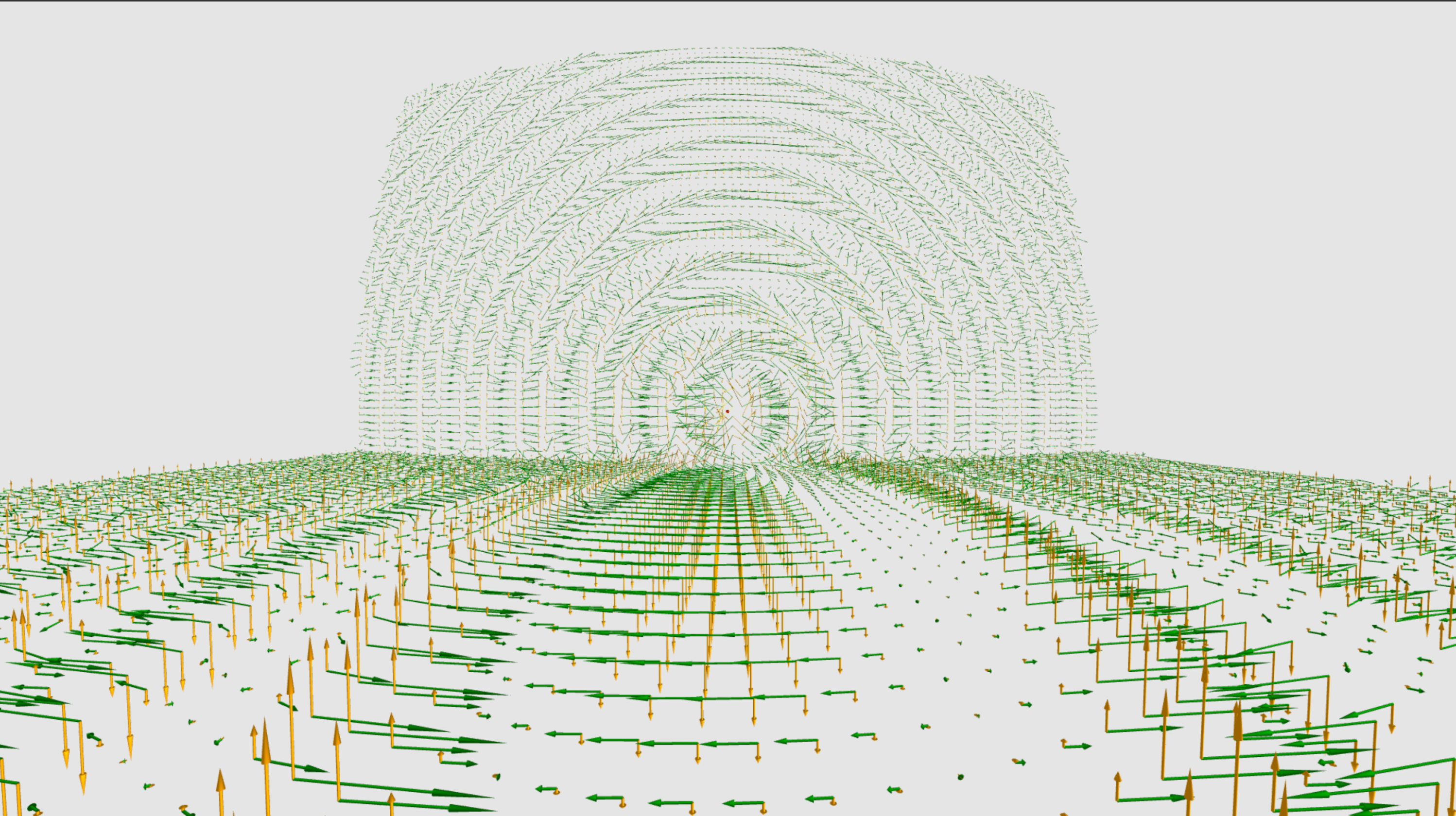}
\caption{
    Scene of the sample program with the Harmonic Oscillator preset, with $c=4\,\tx{grid}/\tx{s}$, Log Reduction Count \(N=1\), and 10-Exponent $n=2$; other explanations of Fig.~\ref{static_charge} apply here too unless otherwise stated. 
    \textbf{Left:} Electromagnetic fields (green and yellow arrows) emitted from the positive point charge (red icosahedron) under forced harmonic oscillation. The emission of electromagnetic waves is apparent both in the vertical and horizontal planes. 
    \textbf{Right:} With the forward acceleration of the player, the planes are Lorentz contracted/elongated as in the Right of Fig.~\ref{static_charge}. On the real-time simulator, one can observe an increase in the oscillation frequency, where the time ``contraction'' of the incoming charge is observed; see Appendix~C in Ref.~\cite{10.1093/ptep/ptx127} for a detailed explanation. 
    In both Left and Right, the horizontal plane exhibits apparently non-spherical waves because it does not correspond to a plane on a fixed time slice of the world frame but on the player's PLC, such that the more distant region corresponds to the further past. 
    \label{oscillating charge fig}
}
\end{figure}

\begin{figure}[tp]
\centering
\includegraphics[width=0.49\textwidth]{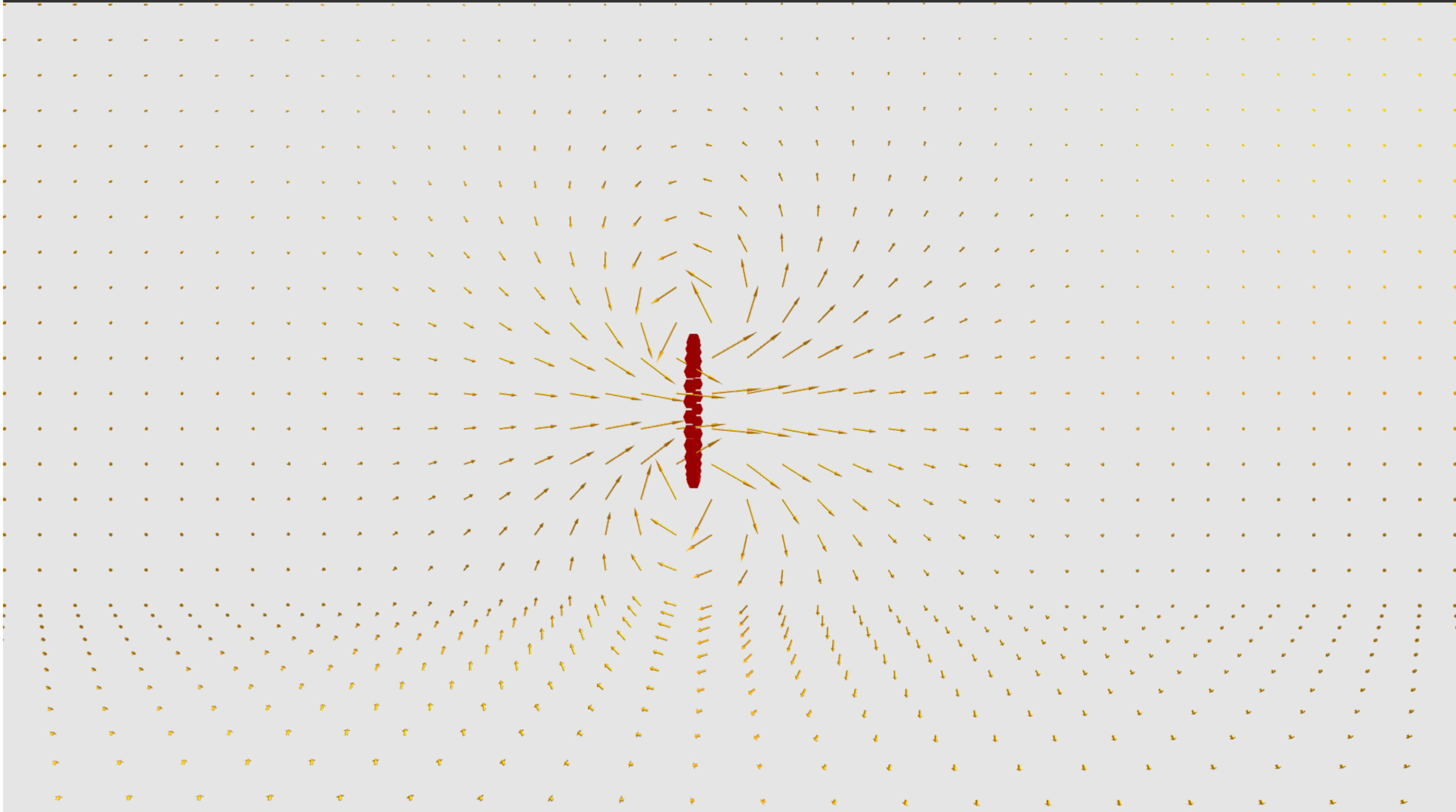}\hfill
\includegraphics[width=0.49\textwidth]{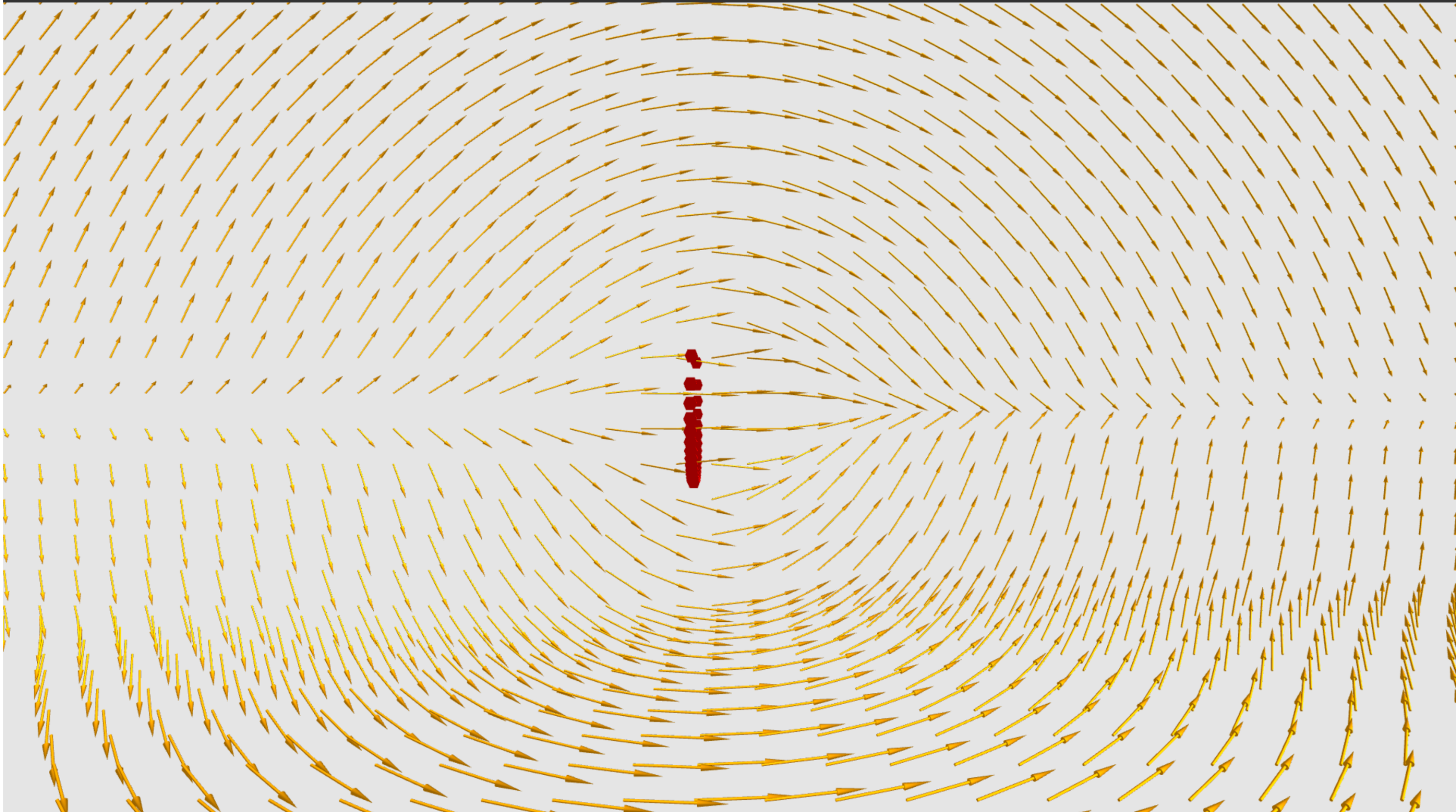}
\caption{
    Scene of the sample program with the Current Loop preset, with \(c=4\,\tx{grid}/\tx{s}\), Log Reduction Count \(N=2\), and 10-Exponent \(n=2\); other explanations from Fig.~\ref{static_charge} apply here as well, unless otherwise stated. The electric field is not displayed here.
    \textbf{Left:} Magnetic field winding around the electric current generated by the rotating loop of positive charges at a non-relativistic speed. The magnetic flux through the loop appears as expected, with the direction of the flux being from left to right inside the loop and from right to left outside the loop.
    \textbf{Right:} At an ultra-relativistic speed of the charges, a vortex-like magnetic field is produced along the rotation axis. Both inside and outside the loop, the direction of the magnetic flux becomes from left to right. The charges on the upper (lower) side move toward (away from) the viewer, becoming sparse (dense) due to Lorentz stretching (contraction), as explained at the end of the caption for Fig.~\ref{static_charge}.
In the real-time simulation, one can observe the dramatic transition from Left to Right. 
    \label{current loop fig}
}

\end{figure}

\subsection{Preset}
Regarding the sources of electromagnetic fields, we offer multiple preset configurations:
\begin{enumerate}
\item Static Neg \& Rotating Pos: A fixed negative charge, paired with a dynamical positive charge.
\item Dynamic Opposite Charges: A pair of positive and negative dynamical charges.
\item Static Charge: A static charge; see Fig.~\ref{static_charge}.
\item Oscillating Charge: A positive charge undergoing predetermined harmonic oscillation; see Fig.~\ref{oscillating charge fig}.
\item Current Loop: A loop of positive charges is initially at rest. Its rotation accelerates towards the speed of light.\footnote{
Specifically, we set the angular velocity \({\df\theta\ov\df t} = \frac{c}{r} \frac{1 + \tanh\fn{b\pn{t-t_0}}}{2}\), where \(r\) is the radius of the loop, \(t_0\) shifts the initial time, and \(b\) is an arbitrary constant. The resultant angle behaves as $\theta\fn{t}=\theta_0+{c\ov2r}\sqbr{t-t_0+{\ln\cosh\fn{b\fn{t-t_0}}\ov b}}$.
}
The electric field is not displayed in the default setup to focus on the magnetic field.
At non-relativistic speeds of the charges, the magnetic field exhibits the typical winding around the current, resulting in flux from left to right inside the loop and from right to left outside it.
As the rotation of the charges accelerates, one can observe a dramatic transition to a vortex-like magnetic flux, with the flux direction being from left to right both inside and outside the loop.
\end{enumerate}
In the first two presets, the dynamical charges adhere to the fully relativistic equation of motion as given in Eq.\eqref{eom simple form}, with their time evolution outlined in Sec.~\ref{time evolution section}.

\subsection{Grid}
For visualizing electromagnetic fields, two rendering options are provided:
\begin{itemize}
\item Visualization only at lattice points located on two mutually perpendicular planes.
\item Visualization at all lattice points within a three-dimensional grid.
\end{itemize}

\subsection{Controls}
On a tablet device, such as a smartphone, acceleration in the direction of sight is achieved by pinching out the screen, while pinching in results in acceleration in the opposite direction. Braking is accomplished by double-tapping the screen. The line of sight can be rotated by swiping.

When using a keyboard, constant acceleration towards the forward, backward, left, and right directions is achieved by pressing the \ttt{W}, \ttt{S}, \ttt{A}, and \ttt{D} keys, respectively. Similarly, upward and downward accelerations are controlled by the \ttt{X} and \ttt{Z} keys, respectively. Rolling is performed using the \ttt{Q} and \ttt{E} keys, while braking is achieved by pressing the \ttt{R} key.

\section{Summary and discussion}\label{Conclusion}

We have presented a fully relativistic and Lorentz-covariant simulation framework for visualizing electromagnetic fields in special relativity. By incorporating the Lorentz transformation, light-cone structure, and the dynamics of charged particles, our approach enables users to directly observe the relativistic intermixing of electric and magnetic fields from the viewpoint of a moving observer.

In our simulation, the electromagnetic field at each spacetime point is determined by the motion of point charges at the intersection of their worldlines with the PLC of that point, ensuring causal consistency and Lorentz covariance. Each charged particle---including the observer---then experiences the Lorentz force from the field evaluated at its own spacetime location. Finally, the observer perceives both the fields and other particles as they appear on their PLC, constructing a consistent visual world from relativistic principles.

A key feature of our simulation is the visualization of observer-dependent electromagnetic fields. Users can witness how electric and magnetic components transform under changes in velocity, thereby gaining intuition for time dilation, length contraction, and the unity of electric and magnetic phenomena. The simulation also illustrates the Lorentz stretching effect---an often-overlooked consequence of field transformations---offering a more comprehensive view of relativistic dynamics.

Beyond its technical contributions, this simulation serves as a pedagogical tool to make abstract concepts in relativity and electromagnetism more accessible. Its interactive nature promotes intuitive understanding and offers potential for use in both classroom and outreach settings.

Future extensions may include incorporating gravitational effects or enhancing the visualization for broader engagement. We hope this work will contribute to both scientific communication and deeper conceptual insight into the relativistic structure of our physical world.

\subsection*{Acknowlegement}
The work of K.O.\ is in part supproted by JSPS KAKENHI Grant No.~23K20855.

\bibliographystyle{JHEP}
\bibliography{refs}

\end{document}